\documentclass{aa}  

\usepackage{graphicx}

\usepackage{bm}
\usepackage{amsmath}	

\usepackage{txfonts}
\usepackage{xcolor}
\usepackage{soul}
\usepackage{natbib}
\bibpunct{(}{)}{;}{a}{}{,} 
\usepackage[colorlinks=true,linkcolor=blue,citecolor=blue,urlcolor=blue]{hyperref}
\begin{document} 
\raggedbottom

   \title{Revisiting thermoelectric effects in the crust of neutron stars}

   \titlerunning{Thermoelectric effects in neutron star crusts}

   \author{Dionysios Gakis \thanks{dgakis@upnet.gr}
          \and
          Konstantinos N. Gourgouliatos \thanks{kngourg@upatras.gr}
          }
          
 \institute{Department of Physics, University of Patras, Patras, Rio, 26504, Greece\\
             }
             
   \date{Received 22 February 2024; accepted 7 July 2024}

 
  \abstract
   {Large thermal variations have been observed in neutron stars that typically are not aligned with density gradients. Such terms may activate the Biermann battery effect, leading to thermoelectric interactions and to the generation of electromotive force.}
   {We aim to identify the possible impact of a temperature anisotropy on the crust of a neutron star can have in the evolution of its magnetic field, through the thermoelectric terms.}
   {We consider a neutron star crust with large temperature gradients, associated with long-lived hot spots, described by a localized Gaussian-type function. We simulate the interplay between the battery term and the Hall and Ohmic evolution numerically for axisymmetric systems.}
   {The results indicate that for crust temperatures of $\sim$$10^9$ K the toroidal field can be amplified up to $\sim$$10^{14}-10^{15}$ G near the points of maximum temperature gradients, and it locally changes the architecture of the poloidal field lines. For internal crustal temperatures of $\sim$$10^8$ K, the temperature gradient generates fields that are lower by about two orders of magnitude. In these cases, saturation is achieved after some hundred thousand years, after which the battery and Ohmic dissipation balance each other, whereas the Hall drift contributes comparatively little to the final field strength, but it can affect its structure.}
   {We conclude that the thermoelectric effect can impact the overall magnetic field evolution, provided that the thermal gradient is maintained for a sufficiently long time. Neutron stars endowned with moderate-strength magnetic fields may be affected by the thermoelectric effect if the hotspots survive for timescales of a few kiloyears. }


   \maketitle
%

\section{Introduction}

Neutron stars are known to sustain the strongest magnetic fields found in the universe, ranging from $10^8$ G in the case of millisecond pulsars \citep[e.g.][]{Possenti:2003} up to $10^{15}$ G for magnetars \citep[e.g.][]{Nakagawa:2009}. The origin of these fields still remains an open question. \cite{Spruit:2008} specifically reviewed the following two scenarios. The field could be inherited by the progenitor main sequence star through flux conservation \citep[e.g.][]{Makarenko:2021}, or it might be generated by dynamo processes before the supernova explosion. The bottom line is that they only appear as plausible explanations for pulsars but not for magnetars; other interesting theories have been suggested for the latter \citep[e.g.][]{Braithwaite:2006,2015Natur.528..376M,2022ApJ...926..111W,2024NatAs...8..298K}.  

Whatever the source is, we have indications that magnetic fields in neutron stars undergo evolution with time. The majority of neutron stars have strong magnetic fields on their surfaces ($\sim$$10^{10}$-$10^{13}$ G), while those with stronger magnetic fields host high-energy phenomena, such flares and bursts, which have been linked to changes of their magnetic field structure \citep[e.g.][]{Mazets:1999, Turolla:2015, Kaspi:2017, Younes:2017, Zelati:2018, Lin:2020, Roberts:2021}. The observed field strengths spanning many orders of magnitude is yet another argument towards the supposed evolution of magnetization in neutron stars given the fact that the older of them are characterized by weaker fields.

Broadly speaking, the structure of a typical neutron star may be divided into its crust and core, which in turn are further split into inner and outer parts. The outer crust is physically described by a degenerate relativistic electron gas flowing on a solid crystal Coulomb lattice of protons and ions, whereas matter consisting of neutron-rich nuclei, at densities between nuclear and the neutron drip point, constitute the inner crust. At greater depth, the outer core is a superconductive superfluid of protons and neutrons with only a few free electrons. In the innermost part of a neutron star, the inner core, the extremely high dominating densities restrict us to only speculations about exotic forms of matter, such as free quarks or other unknown particles \citep[e.g.][]{Baym:2018}. The part of a neutron star between the crust and the magnetosphere (governed by plasma) is the atmosphere, containing the ocean (sometimes called envelope). Although it might be possible that the magnetic field permeates both crust and core of a neutron star, simulations typically only considered the crustal magnetic field, and only a handful of recent studies included the core as well \citep[e.g.][]{Ciolfi:2013,Moraga:2023,Skiathas:2024}. In our paper, we also assume that the magnetic flux in the core is frozen, and we study its evolution in the crust.

There are typically three processes governing the magnetic field evolution in neutron stars \citep{Goldreich:1992}. The Hall drift is the transport of magnetic flux by a moving electron fluid rising electric currents and is conservative. The conversion of magnetic energy into heat because of the crust's finite conductivity is called Ohmic dissipation. Third, the ambipolar diffusion, another dissipative process, is caused by the relative movement of the magnetic field and charged particles to neutrons \citep[e.g.][]{Shalybkov:1995}. The latter effect is only important in the core, where neutrons are abundant, and hence is disregarded in the context of this study. The inner battle between Hall and Ohmic terms in neutron star crusts has been thoroughly analyzed in the literature \citep[e.g.][]{Hollerbach:2002,Cumming:2004,Pons:2007,Kojima:2012,Gourgouliatos:2014,Gourgouliatos:2016}. However, the magnetic and thermal interactions cannot be considered isolated in neutron star physics, and because of their nonlinear interactions, it has only recently become possible to simulate
their coupling in 2D \citep[e.g.][]{Vigano:2013} and 3D \citep[e.g.][]{Igoshev:2021,de-Grandis:2020,Dehman:2023,Ascenzi:2024}. Motivated by these efforts, in this paper, we examine a yet another possible source of magnetic field production, driven the Biermann battery mechanism \citep[]{Biermann:1950}.

The possible generation of magnetic field through thermoelectric instabilities was initially studied in laboratory environments for laser-heated plasmas \citep[e.g.][]{Stamper:1971,Tidman:1974}. The Biermann battery mechanism has been proposed as a mechanism for the formation of magnetic structures by weak seed fields in astrophysics contexts as well, for instance, on cosmological scales \citep[e.g.][]{Gnedin:2000}. Later works motivated by \cite{Biermann:1950} \citep[e.g.][]{Roxburgh:1966} suggested that differential rotation in fluid stars could produce a misalignment in isobars and equipotentials and consequently, battery effect. This effect is generally rather small compared to the dynamo action by helical motion in a fluid which is far more effective. Nonetheless, thermoelectric effects might be important in neutron stars because dynamo action cannot occur in solid crusts.

The thermoelectric mechanism was first studied in the cores of white dwarfs by \cite{Dolginov:1980a,Dolginov:1980b}. However, as noted by \cite{Blandford:1983}, the mechanism should not be as relevant in white dwarfs as their crystallization occurs when they are very old and little thermal energy remains. Thus, neutron stars remain as main candidates for this phenomenon.

In neutron star crusts, the battery mechanism could work the following way. For some reason, it is possible that thermal anisotropies are formed within the star. Such temperature gradients, along with density gradients that appear naturally within depth, may circulate the electron fluid, carrying both thermal and electric currents. This thermoelectrically generated electric field acts as a battery, generating a magnetic field perpendicular to the density and temperature gradients.

The first attempt of studying such effects in neutron star crusts was by \cite{Blandford:1983} who extended the work of \cite{Urpin:1980} about thermogalvanomagnetic effects in degenerate stars\footnote{We note that in a few cases in the literature, the induction of electromotive force from thermal gradients is referred to as "thermomagnetic" or "thermogalvanomagnetic". In this work we refer to this as the thermoelectric effect, as the term "thermomagnetic" is related to the change of the magnetization of material due to temperature variations, and this use may lead to confusion.} In particular, they examined two mechanisms for generation of magnetic flux, in the solid crust and in the melted envelope layer above it. For the first case, they showed that heat carried by degenerate electrons could cause thermoelectric instability in the solid crust, resulting in horizontal magnetic fields that grow exponentially with time under appropriate conditions. In the second case, they found that since pressure and density gradients must be parallel, the battery mechanism cannot work in the same way as previously. However, the liquid layers contain horizontal magnetic fields as well, heat flux can drive circulation that amplifies the field strength (possibly leading to a dynamo effect), and hence, flux may be convected from the liquid to the solid layers. If these instabilities develop, the field rapidly rises to $\sim$$10^{12}$ G, with additional growth saturating at $\sim$$10^{14}$ G, the point at which non-linear perturbations describing the field evolution must be included, or when the magnetic and the lattice yield stresses are comparable.

Some years later, \cite{Urpin:1986} considered unstable modes in shallower depths, consisting of free degenerate electrons and fully ionized ions in a liquid or gaseous state, and unlike \cite{Blandford:1983}, they did not include hydrostatic equlibrium in their considerations. They worked in the linear frame as well and showed that thermoelectric modes are unstable if the star rotates. They were able to deduce toroidal field growth in quite hot neutron stars in a very narrow region close to surface. The growth rate was found to depend on the surface temperature and the typical size of the field fluctuation (with a maximum at $\sim$100 m). Therefore, the generated field has a strong non-dipolar character, as large-scale fields decay in linear approximation. Conversely, \cite{Blondin:1986} suggested that the accretion process in young neutron stars may remove some of their magnetic field in a short period of time.

Next, the amplification of neutron star magnetic fields through thermoelectric effects was further analyzed extensively in a series of six papers \citep[]{I,II,III,IV,V,VI}. In the first paper \citep[]{I}, the general formalism was presented, which considered an isolated non-accreting slowly rotating (thus neglecting deviations from spherical symmetry) neutron star and examining the field growth in the limit in which the Lorentz force can be safely ignored and magnetic pressure remains negligible compared to other forms of pressure (i.e. the magnetic field can alter the thermal but not the density profile of the star). Thermoelectric amplification was sought in the liquid layer, but as opposed to \cite{Urpin:1986}, the outer boundary condition was slightly different, as were the mathematical methods for simultaneously solving the system of heat transport and induction equations.  

In the linear approximation \cite[]{II}, which is valid for moderate field strengths ($\lesssim 10^{11}$ G), the growth and decay rates of the magnetic field components for different multipolar components were calculated for several surface temperatures and gravitation intensities. The authors found that small-scale ($n$ $\sim$ 1000) toroidal components grow exponentially in some years under certain conditions within a thin layer ($\sim$20-100 m) below the surface creating sinusoidal variations with meridional wavelengths of $\sim$100 m (like the results obtained by \cite{Urpin:1986}). Large-scale modes such as dipole and quadruple slowly decay in linear approximation. These outputs were reinforced by the analytical approximations performed in \cite{VI}. In this last paper, simple formulae for the maximum position and width of the temperature and field functions (assumed to be Gaussian) were derived. Interestingly, these two functions were found to be quite symmetric despite the strong dependence of the parameters on density. It was shown that the layer and maximum position of the functions loosely depend on $n$, but with larger $n$, the thickness decreases and the maximum position is shifted outwards. The authors were finally able to reproduce the described behavior of the the growth rate in their analytical formula and reached a straightforward criterion for the occurrence of thermoelectric instability.

In the other papers, non-linear calculations were performed. \cite{III} investigated the dominating non-linear terms on the toroidal small-scale magnetic field rapid growth, and yielded that the rate decreases with increasing field strength because of some form of self-coupling. Asymptotically, a stationary solution at $10^{11}$ G was obtained. Saturation occurs when the characteristic time (given by the inverse of electron Larmor frequency) becomes comparable with the relaxation time for electron-ion collisions (in principle, when the electrons are forced to spiral around the field lines, and hence, collisions become improbable). Since the seed field is unknown a priori, \cite{IV} introduced averaged small-scale modes to obtain the structure of an individual axisymmetric mode, the relative weights of which are defined by initial conditions (assuming chaotic distributions). Regardless of the seed field geometry, the axisymmetric toroidal large-scale modes with even multipolarity in particular are strongly induced, while all other modes (including poloidal ones) are amplified at the same rate, and their (absolute) field values are significantly lower. Additionally, in contrast to the linear approximation, large-scale modes can be amplified rapidly in non-linear coupling with small-scale modes, and some selection rules of this coupling were identified. \cite{VI} described the evolution of the strongly induced axisymmetric large-scale toroidal field modes with even multipolarities (without considering their coupling with the poloidal field). Their evolution seems rather insensitive to the underlying star properties, and they may be split into a phase of exponential growth followed by inward diffusion along with gradually slowing down further enhancement.

In spite of these advancements and the somewhat promising results, thermoelectric effects on neutron stars have been quite overlooked in recent years. Some encouragements to explore this direction have indeed been formulated (for instance, we refer an interested reader to \cite{Geppert:2017} for a review of the magneto-thermal evolution of neutron stars) along with individual speculations about their application \citep[e.g.][]{Konenkov:2001,de-Grandis:2020,Geppert:2021}. Nevertheless, a thorough numerical investigation of the battery term and its battle with the Hall and Ohmic effects has not been done. This is the goal of the current study.

On the grounds of a comprehensive exploration of the possible effects of a thermoelectric battery term on the magnetic field evolution within a neutron star crust, we have performed a number of numerical simulations solving the magnetic induction equation for different sets of the stellar configurations. We aim to identify under which conditions the battery mechanism can first be activated, and then prevail over the other two processes. Particularly, we are interested in the range of temperatures and initial magnetic field structures and intensities required to give some predictable results, as well as in the typical timescales for the relevant phenomena to evolve. Finally, we assess the extent to which the ensemble of parameters we examined may lead to conditions that can be applied to realistic neutron stars.

This paper is organized as follows. Section \ref{sec:sec2} contains the set of the underlying equations describing the problem and in Section \ref{sec:sec3}, we discuss the numerical approach implemented to solve them. We continue with analyzing the results from the simulations (Section \ref{sec:sec4}) and discuss their implications for realistic neutron stars in Section \ref{sec:sec5}. A summary of this work is given in Section \ref{sec:sec6}.


\section{Mathematical setup} \label{sec:sec2}


\subsection{Magnetic induction equation}

We work in the one-fluid approximation of the crust, in which ions are considered to be static and to form a crystal Coulomb lattice, and in which electrons have the freedom to move. They are thus the carriers of electric current and thermal energy as the phonon contribution is considered negligible if the magnetic field is strong enough, stronger than $\sim$10$^{12}$ G, as deduced by \cite{Chugunov:2007}. We remark though that for such magnetic fields the phonon conductivity and the electron conductivity across the magnetic field are comparable, and this would enforce an anisotropic electron conductivity; nevertheless thermal conductivity is still assumed to be isotropic in this work. The typical electron velocities are low enough to resort to non-relativistic electron magneto-hydrodynamics (e-MHD). Thus, the electric current density $\bm{j}$ is given by:
\begin{equation}
\bm{j} = - e n_e \bm{\upsilon},
\end{equation}
where $-e$ is the elementary electron charge $n_e$ is the electron number density, and $\bm{\upsilon}$ is the mean electron velocity.

In the case of finite electric conductivity $\sigma$, the electric field $\bm{E}$ may be written through the generalized Ohm's law:
\begin{equation}\label{2}
    \bm{E}= - \frac{\bm{\upsilon} \times \bm{B}}{c} + \frac{\bm{j}}{\sigma} - \frac{1}{e} S_e \nabla T,
\end{equation}
in which $c$ is the speed of light, $\bm{B}$ is the magnetic field, $T$ is the temperature and $S_e$ represents the electron entropy, defined as:
\begin{equation}
    S_e = \left(\frac{\pi^4}{3 n_e} \right)^{1/3} \frac{k_B^2 T}{c \hbar},
\end{equation}
where $k_B$ is Boltzmann's constant, and $\hbar$ is the reduced Planck constant. In this paper, the last term of eq.~(\ref{2}) is of particular importance. It expresses an additional component of the electric field, the component that is formed by a temperature gradient. In its general form, this electric field component is given by:
\begin{equation}
    \bm{E} = \hat{G} \nabla T,
\end{equation}
where the tensor $\hat{G}$ is the thermopower and has two parts: an isotropic part (Seebeck term) and an anisotropic part (Ettingshausen-Nernst term). The Biermann battery is due to the isotropic term \citep{de-Grandis:2020}. This term has the form  $G_{ij} = -S_e/e \delta_{ij}$ for a completely degenerate Fermi gas \citep{Gourgouliatos:2022}, as written above.

The magnetic fields have currents as sources, according to Amp\`ere's law:
\begin{equation}
    \bm{j}= \frac{c}{4 \pi} \nabla \times \bm{B}.
\end{equation}
Faraday's law states that any change of the magnetic field induces an electric field as: 
\begin{equation}
    \frac{\partial \bm{B}}{\partial t} = -c \nabla \times \bm{E}.
\end{equation}
Combining the above equations, we obtain the full magnetic induction equation:
\begin{eqnarray}\label{1}
\frac{\partial \bm{B}}{\partial t} = -c\nabla \times  \left(\frac{1}{4 \pi e n_e} (\nabla \times \bm{B}) \times \bm{B} + \frac{c}{4 \pi \sigma} \nabla \times \bm{B} - \frac{1}{e} S_e \nabla T \right)\,
\end{eqnarray}
In this equation, the first term expresses the Hall effect, which is dependent on $\bm{B}^2$, and the second term refers to Ohmic dissipation \citep{Jones:1988,Goldreich:1992}. The third term is the thermal battery, which is independent of $\bm{B}$, and its strength is modulated by the $T$ gradient and the thermopower.

We now closely follow the formalism by \cite{Gourgouliatos:2014}. In axisymmetry, we decomposed the magnetic field into a poloidal ($\bm{B}_P$) and a toroidal ($\bm{B}_T$) component, and traced their evolution in terms of two scalar quantities, $\Psi(r, \theta)$ and $I(r, \theta)$:
\begin{equation}
\bm{B} = \bm{B}_P + \bm{B}_T = \nabla \Psi \times \nabla \phi + I \nabla \phi\,,
\end{equation}
where we use spherical coordinates ($r$, $\theta$, $\phi$) and $\nabla \phi = \hat{\phi}/r sin\theta$. By construction, the above expression fulfills the Gauss law for the magnetic field, $\nabla \cdot \bm{B} = 0$, provided that $\Psi$ and $I$ are differentiable quantities. The function $\Psi$ is referred to as the flux function (identical in form to the Stokes stream function), since $2 \pi \Psi$ is the poloidal magnetic flux passing through a spherical cap of radius $r$ and opening angle $\theta$. $I$ is the poloidal current function, as $c I/2$ expresses the poloidal current passing through the same spherical cap. The magnetic field lines lie on the surfaces on which $\Psi$ is constant, while the current or electron flow lines lie on surfaces of constant $I$.

We now express eq.~(\ref{1}) in terms of the quantities $\Psi$ and $I$. As in \cite{Reisenegger:2007}, we define the quantity 
\begin{equation}
\chi={\frac{c} {4\pi {\rm e}n_{\rm e} r^{2}\sin^{2}\theta}}\,,
\end{equation}
the toroidal current as
\begin{equation}
\bm{j}_T={c\over 4\pi}\nabla\times\bm{B}_{P}=-{c\over 4 \pi} \Delta^{*}\Psi \nabla \phi\,,
\end{equation}
and the angular velocity of the electrons is
\begin{equation}
\Omega= -{j_T\over n_{\rm e} {\rm e} r\sin\theta}=\chi \Delta^{*} \Psi\,,
\end{equation}
where the Grad-Shafranov operator is given by
\begin{equation}
\Delta^{*}=\frac{\partial^{2}}{\partial r^{2}} +\frac{\sin\theta}{r^{2}}\frac{\partial}{\partial \theta}\left(\frac{1}{\sin\theta}\frac{\partial}{\partial \theta}\right).
\end{equation}
After some vector calculus manipulation, the coupled differential equations for $\Psi$ and $I$ are as follows. The equation related to poloidal field evolution is
\begin{eqnarray}\label{13}
\frac{\partial \Psi}{\partial t} -r^{2}\sin^{2}\theta\chi (\nabla I \times \nabla \phi)\cdot \nabla \Psi =\frac{c^{2}}{4 \pi \sigma}\Delta^{*}\Psi\,,
\label{dPSI}
\end{eqnarray}
and the equation concerning toroidal field evolution is
\begin{eqnarray}\label{14}
\frac{\partial I}{\partial t} +r^{2}\sin^{2}\theta (\left( \nabla \Omega \times \nabla \phi \right) \cdot \nabla \Psi + I \left(\nabla \chi \times \nabla \phi \right) \cdot \nabla I ) \nonumber\\= \frac{c^{2}}{4 \pi \sigma}\left(\Delta^{*}I-\frac{1}{\sigma}\nabla I \cdot \nabla \sigma\right) + \frac{c}{e} r\sin\theta( \nabla S_e \times \nabla T ) \cdot \hat{\phi} \,.
\label{dI}
\end{eqnarray}
hence, the toroidal and poloidal components of the field are strongly coupled to each other in their evolution.

The poloidal and toroidal energies are computed as follows:
\begin{equation}
    E_{pol}=\frac{1}{8 \pi} \int_V (B_r^2+B_{\theta}^2)\,dV,
\end{equation}
\begin{equation}
    E_{tor}=\frac{1}{8 \pi} \int_V B_{\phi}^2\,dV,
\end{equation}
where $B_r=-r^{-2} \partial \Psi/\partial \mu$, $B_{\theta}=-1/(r sin\theta)  \partial \Psi/\partial r$ and $B_{\phi}=I/(r sin\theta)$ are the components of the magnetic field in spherical coordinates.

\subsection{Crust properties}

We now outline our model of the microphysical parameters within the crust, used to solve eqs.~(\ref{13}) and~(\ref{14}). First, the crust is modeled as a spherical shell, covering the outermost 5\% part of the star in terms of radius. Hence, the crust extends from the crust-core boundary, $r_{cc}$, up to the base of the ocean, $r_{out}$, and $r_{cc}/r_{out}=0.95$. As the radius of our modeled star is 11.3 km, we only consider the last $\sim$500 m of it.

We assume that the density throughout the crust varies as a function of depth, approximating the formula deduced by \cite{Chamel:2008} (adopted by \cite{Karageorgopoulos:2019} as well) with the following expression:
\begin{equation}
   {\rho}(r) = \left(1 + \left(\frac{r_{out}-r}{r_{out}-r_{cc}}\right)^4 \frac{\rho_{cc}}{\rho_{out}} \right) \rho_{out},
\end{equation}
where $\rho_{cc} = 1.3 \times 10^{14}$ g cm$^{-3}$ is the density at the base of the crust and $\rho_{out}$ is the density of the outermost point considered (bottom of the ocean). We note that $\rho_{out} \ll \rho_{cc}$, thus the formula is sufficiently accurate at $r=r_{cc}$ (precisely ${\rho}(r=r_{cc}) = \rho_{cc}+\rho_{out}$).

Crystallization occurs when the electrostatic potential energy between nuclei dominates over thermal energy. Their ratio $\Gamma$ is the so-called Coulomb parameter:
\begin{equation}\label{Gamma}
    \Gamma = \frac{Z^2 e^2}{a_I k_B T}.
\end{equation}
The critical value of $\Gamma$ is 175, which corresponds to the melting point, i.e. crystallization occurs for larger values \citep{Potekhin:2000}. We assume a pure iron crust ($Z=26$, $A=56$). The ion sphere radius (also referred to in literature as interionic spacing) is: 
\begin{equation}
    a_I = \left (\frac{4 \pi n_I}{3} \right)^{-1/3}.
\end{equation}
The density $\rho_{out}$ at the bottom of the ocean may be found using the numerical density $n_I$ (for a fully ionized crust):
\begin{equation}
    \rho_{out} = n_{I} m A,
\end{equation}
which by construction is a function of temperature, as is evident from eq.~(\ref{Gamma}). As explained in Section \ref{sec:sec3.2}, in our model we use a meridionally varying temperature, however, in order to determine $\rho_{out}$ corresponding to the melting point, we use the base temperature $T_0$ without the fluctuation. The liquid part of a neutron star (ocean) exists for lower densities, but is not included in our considerations due to its comparably small size ($\sim$ some cm) and the fact that e-MHD fails in this region. 

In general, the electron number density $n_e$ is given by:
\begin{equation}
    n_e = \frac{Z}{A} (1 - x_{fn}) \frac{\rho}{m_b},
\end{equation}
in which $x_{fn}$ is the fraction of free neutrons outside ions and $m_b$ is the atomic mass unit. \cite{Lander:2019} performed a polynomial fitting of $Z$, $A$, and $x_{fn}$ in terms of $\rho$ utilizing the equation of state by \cite{Douchin:2001}, and their result may be approximated accurately enough by the following straightforward expression: 
\begin{equation}
    n_e = 10^{36} (1.5 \tilde{\rho}^{2/3} + 1.9 \tilde{\rho}^{2}) \ cm^{-3}, 
\end{equation}
which was also used later by \cite{Gourgouliatos:2021}. The scaled density is $\tilde{\rho} = \rho/\rho_{cc}$.

Regarding conductivity, we assume that it scales as $\sigma \propto n_e^{2/3}$, based on phonon scattering or impurity scattering \citep{Cumming:2004}, being  $\sigma (r = r_{cc} )= 10^{24} s^{-1}$ at the crust-core boundary. It is not considered a function of temperature in order to keep our code manageable and focus on the physics of the magnetic field evolution.

The typical timescale for the Ohmic evolution is:
\begin{equation}
t_{\rm Ohm} \sim {4\pi \sigma L^2\over c^2} = 566\ {\rm Myr}\ \left({L\over 11.3\ {\rm km}}\right)^2\left({\sigma\over 10^{24}\ {\rm s^{-1}}}\right),
\end{equation}
whereas the Hall timescale is
\begin{equation}
t_{\rm Hall} \sim {4\pi {\rm e} L^{2} n_{\rm e} \over cB} = {81.2\ {\rm Myr}\over B_{14}}\ \left({L\over 11.3\ {\rm km}}\right)^2\left({n_{\rm e}\over 10^{36}\ {\rm cm^{-3}}}\right),
\end{equation}
and the timescale of the battery is
\begin{equation} \label{23}
t_{\rm battery} \sim {e L^2 B \over c T S_e} = 33.6\ {\rm Myr} {B_{14}\over T_9^2}\ \left({L\over 11.3\ {\rm km}}\right)^2\left({n_{\rm e}\over 10^{36}\ {\rm cm^{-3}}}\right)^{1/3},
\end{equation}
where $B_{14}$ is the strength of the magnetic field expressed in units of $10^{14}$ G and $T_9$ is the temperature in $10^9$ K. The relative importance of the Hall and Ohm terms is gauged by the ratio of these timescales, which is called magnetic Reynolds number (or Hall parameter):
\begin{equation}
R_M = \frac{t_{\rm Ohm}}{t_{\rm Hall}} = \frac{\sigma |B|}{c e n_e}.
\end{equation}
$R_M$ does vary through the crust and time, and in our simulations we keep it a few times of 10$^2$ at most, because when it reaches some 10$^3$ simulations become unstable. The corresponding dimensionless parameter for the battery term is expressed as:
\begin{equation}
R_T = \frac{t_{\rm battery}}{t_{\rm Hall}} = \frac{3^{1/3} c \hbar |B|^2}{4 \pi^{7/3} k_B^2 n_e^{2/3} T^2},
\end{equation}
and ranges at values between $\sim$10$^{-2}$ and 10$^3$ in our simulations.

\section{Numerical Setup} \label{sec:sec3}

\subsection{Numerical implementation} \label{sec:sec3.1}

Our simulations are performed in a Fortran 90 environment. We consider the magnetic field of the star as axisymmetric, and our calculations are therefore restricted to 2D. We discretized the numerical domain into $r$ and $\mu =\cos \theta$ (in order to have a better control of the grid). The resolution is typically $100 \times 100$ but we experimented in several cases with a double analysis to confirm the numerical convergence of our outputs. 

Derivatives are numerically calculated using finite differences; a central difference scheme is used to calculate spatial derivatives, and a three- and five-point stencil for second and third derivatives, respectively. The numerical method adopted for the time intergations of eqs.~(\ref{13}) and~(\ref{14}) is the second order Adams-Bashford method (AB2). The timestep is adaptable, according to the Courant condition \citep{Courant:1952}:
\begin{equation}
    \Delta t  \leq \frac{\Delta r}{\upsilon_{max}}
\end{equation}
in which $\Delta r$ is the grid spacing and we set as $\upsilon_{max}$ the maximum electron plasma velocity, as we work in e-MHD.

Three sets of boundary conditions are adopted, enclosing the numerical domain of our simulation world. Regarding the surface, we assume a vacuum magnetic field outside of star using a multipole fit. This implies that $B_\phi=0$ at the surface ($r=r_{out}$), giving a multipolar current-free poloidal magnetic field, $I(r_{out},\theta) = 0$, so that no current flows from the star to magnetosphere. The continuity of $B_\theta$ comes as follows. In the absence of current outside of the star, we may write $\nabla \times \bm{B} = 0$, and hence, the vacuum field can be expressed as $\bm{B}=\nabla V$, where $V$ is a scalar function. Along with $\nabla \cdot \bm{B} = 0$, we get Laplace equation, $\nabla^2 V = 0$. Working in axisymmetry, we decompose the external field into a sum of multipolar expansion and obtain \citep{Marchant:2011}:
\begin{equation}
    V(r, \theta)= \sum_{l=1}^{\infty} \frac{a_l}{r^{l+1}} P_{l}(\cos \theta),
\end{equation}
where $P_{l}(\cos \theta)$ are the Legendre polynomials. We have:
\begin{equation}
    B_r(r, \theta)= \frac{\partial V}{\partial r} = - \sum_{l=1}^{\infty} (l + 1) \frac{ a_l}{r^{l+2}} P_{l}(\cos \theta),
\end{equation}
and hence the coefficients are calculated as:
\begin{equation}
    a_l = -r^{l+2}\frac{2l + 1}{2l + 2}
\int_{0}^{\pi} P_l(\cos \theta)  B_r(r, \theta) \sin \theta \, d\theta.
\end{equation}
Thus, $B_{\theta}$ is given by:
\begin{equation}
   B_{\theta}(r, \theta) = \frac{1}{r} \frac{\partial V}{\partial \theta} = \sum_{l=1}^{\infty} \frac{a_l}{r^{l+2}} \frac{dP_{l}(\cos \theta)}{d\theta} = -\frac{1}{r \sin \theta} \frac{\partial \Psi}{\partial r},
\end{equation}
giving the boundary condition:
\begin{equation}
   \frac{\partial \Psi (r_{out}, \theta)}{\partial r} = -\sin \theta \sum_{l=1}^{\infty} \frac{a_l}{r_{out}^{l+1}} \frac{dP_{l}(\cos \theta)}{d\theta}.
\end{equation}

Field lines are not allowed to cross the symmetry axis, that is the functions $\Psi$ and $I$ functions are set to zero on the axis of the star, i.e. $\Psi(r, 0) = \Psi(r, \pi) = I(r, 0) = I(r, \pi) = 0$. Concerning the inner boundary, $\partial I/\partial r = 0$, whereas we consider two cases for the boundary condition for $\Psi$. For the magnetic field confined in the crust, it is $\Psi(r_{cc}, \theta) = 0$ or equal to a time-independent function $f$, $\Psi(r_{cc}, \theta) = f (\theta)$, when allowed to permeate to the core. The crust-confined case is motivated by a magnetic field expelled out of the core faster than thermal or magnetic timescales due to the Meissner effect assumption \citep[e.g.][]{Lander:2014, Passamonti:2017}. In the core-threading case, no magnetic field evolution is allowed in the core itself or at the crust-core boundary, because the evolution timescales of the magnetic field in the core are typically much longer than those of the crust.

\subsection{Initial conditions} \label{sec:sec3.2}

We examine two scenarios for the initial poloidal magnetic field. In the option in which the field is restricted in the crust, the initial poloidal flux is given by the expression:
\begin{equation} \label{26}
    \Psi_1  (t=0)= \Psi_0 \left(1.05 - \frac{r}{r_{out}} \right) \left(\frac{r}{r_{out}} -  \frac{r_{cc}}{r_{out}} \right) sin^2\theta,
\end{equation}
while it is slightly different in the alternative in which the field is assumed to thread the core (without evolving within it, however, as explained in Section \ref{sec:sec3.1}):
\begin{equation} \label{27}
    \Psi_2  (t=0)= \Psi_0 \left(1.05 - \frac{r}{r_{out}} \right) \frac{r}{r_{out}}  sin^2\theta.
\end{equation}
In both situations, the intensity of the field is controlled by the parameter $\Psi_0$ (in units of $6.4 \times 10^{24}$ G cm$^2$). The poloidal current function is fixed at zero at the beginning of the simulations ($I_0=0$). 

The temperature profile of the neutron star is of crucial importance for our calculations too. We choose to adopt a Gaussian-type function for the temperature distribution:
\begin{equation} \label{28}
    T(r,\theta) = T_0 + \lambda T_0 \ exp \left( -\frac{(r-r_0)^2}{\sigma_r^2}\right) exp \left(-\frac{(\mu-\mu_0)^2}{\sigma_{\mu}^2} \right),
\end{equation}
which spreads over both $r$ and $\mu$ (with regard to the induction equation, the meridional gradient is mostly relevant). In this formula, a hot region centered at the point $(r_0,\mu_0)$ with a spread of $1\sigma$ (drop to a factor of $1/\mathrm{e}$) controlled by the parameters $\sigma_r$ and $\sigma_{\mu}$ in radial and meridional directions, respectively, is assumed to be superimposed on a background of constant temperature $T_0$ ($r_0$ and $\sigma_r$ are normalized over the simulated radius, i.e. they are expressed in units of $r_{out}$). The magnitude of the additional temperature at the point $(r_0,\mu_0)$ is $\lambda$ times $T_0$, where $\lambda$ can be any positive number determining the steepness of the temperature gradient.

We are of course aware that in most types of stars, their temperature should rise moving radially inward to their core. However, since our model considers only the outermost 5\% of the neutron star (its crust), it is rather safe to assume that a uniform background temperature distribution governs this region (though it is true that any temperature gradient is rather confined precisely to the outermost layers). The additional heat region was added to account for any temperature anisotropy that might rise in the outer layers of the neutron star (the reasons for this behavior are explained more thoroughly in Section \ref{sec:sec5}), which is subsequently supposed to spread throughout the star. As for the choice of the temperature function (Gaussian profile), lacking any essential evidence of its mathematical formulation, we argue that exponential decays appear rather naturally in such environments. Besides, in any tests we performed with other functions with similar profiles (e.g., quadratic) we found no signs of significant changes in our results.

\begin{figure} 
\includegraphics[width=0.45\textwidth]{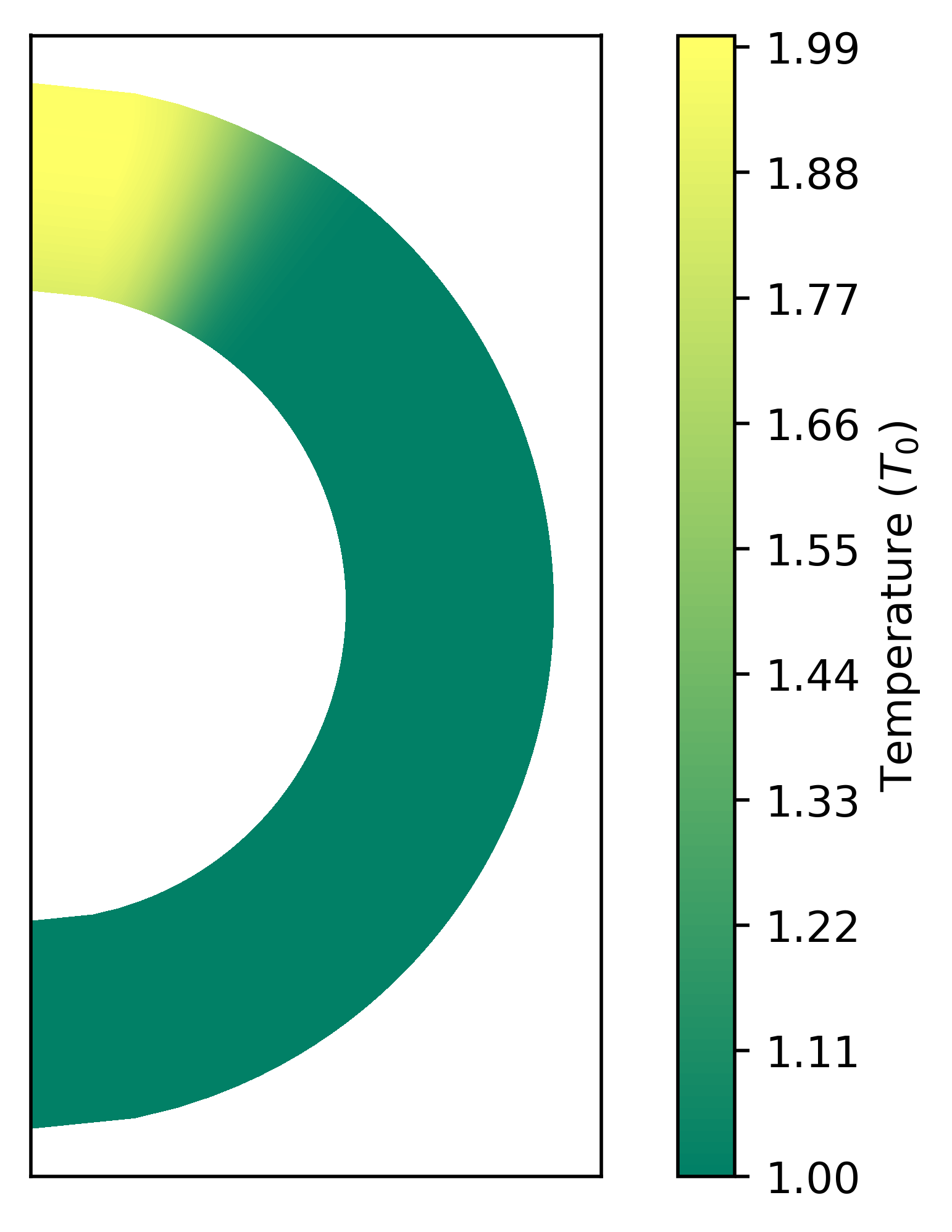}
\caption{Temperature profile for $\lambda=1$, $r_0=0.99$, $\mu_0=0.99$, $\sigma_r=0.1$ and $\sigma_{\mu}=0.1$.}
\label{fig:1}
\end{figure}

We are mostly interested in the influence of the temperature gradient on the magnetic field architecture, hence the point of interest is therefore mainly limited to a region near the pole, in which the temperature is highest. For most of our simulations, we set the central point equal to ($r_0=0.99, \mu_0=0.99$) 
and the standard deviations $\sigma_r=0.1$ and $\sigma_{\mu}=0.1$, while unless otherwise stated we keep $\lambda=1$ (i.e. the highest temperature is twice the background temperature). Sections \ref{sec:sec4.1} and \ref{sec:sec4.2} are dedicated to exploring the field evolution with different combinations of $T_0$ (high and low, accordingly) and $\Psi_0$. A representative temperature profile is presented in Fig.~\ref{fig:1} (the star's crust has been scaled up by a factor of eight for visualization purposes in all such figures in this paper). 
On the other hand, alternative values for the remaining parameters in eq.~(\ref{28}) ($\lambda$, $r_0$, $\mu_0$, $\sigma_r$, $\sigma_{\mu}$) are examined in Section \ref{sec:sec4.3}. 

Lastly, it should be remarked that throughout the simulation time-span, the temperature is kept constant in every run. In other words, we do not account for the evolution of temperature with time and assume that some form of heat source maintains it for some reason at the same levels. While this is apparently an over-optimistic assumption (especially for the timescales considered), it is a necessary approximation we need to make at the current time, in order to examine battery effects in neutron stars and to keep our calculations manageable. Future studies should include the thermal and magnetic interactions in a more realistic context. 

\section{Results} \label{sec:sec4}

\begin{figure*}
\begin{tabular}{p{0.3\textwidth} p{0.3\textwidth} p{0.3\textwidth}}
     a\includegraphics[width=0.312\textwidth]{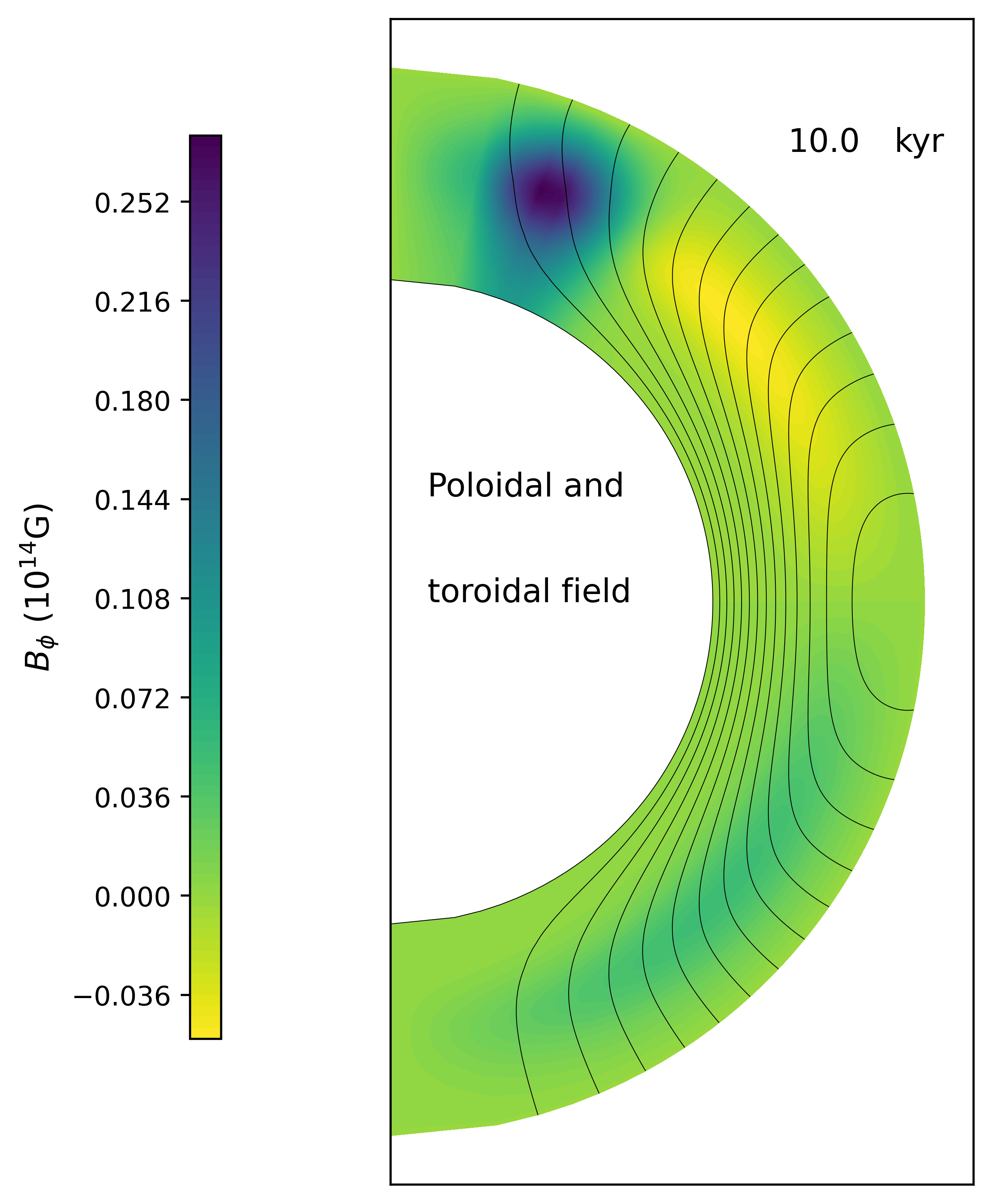}&
     ~b\includegraphics[width=0.3\textwidth]{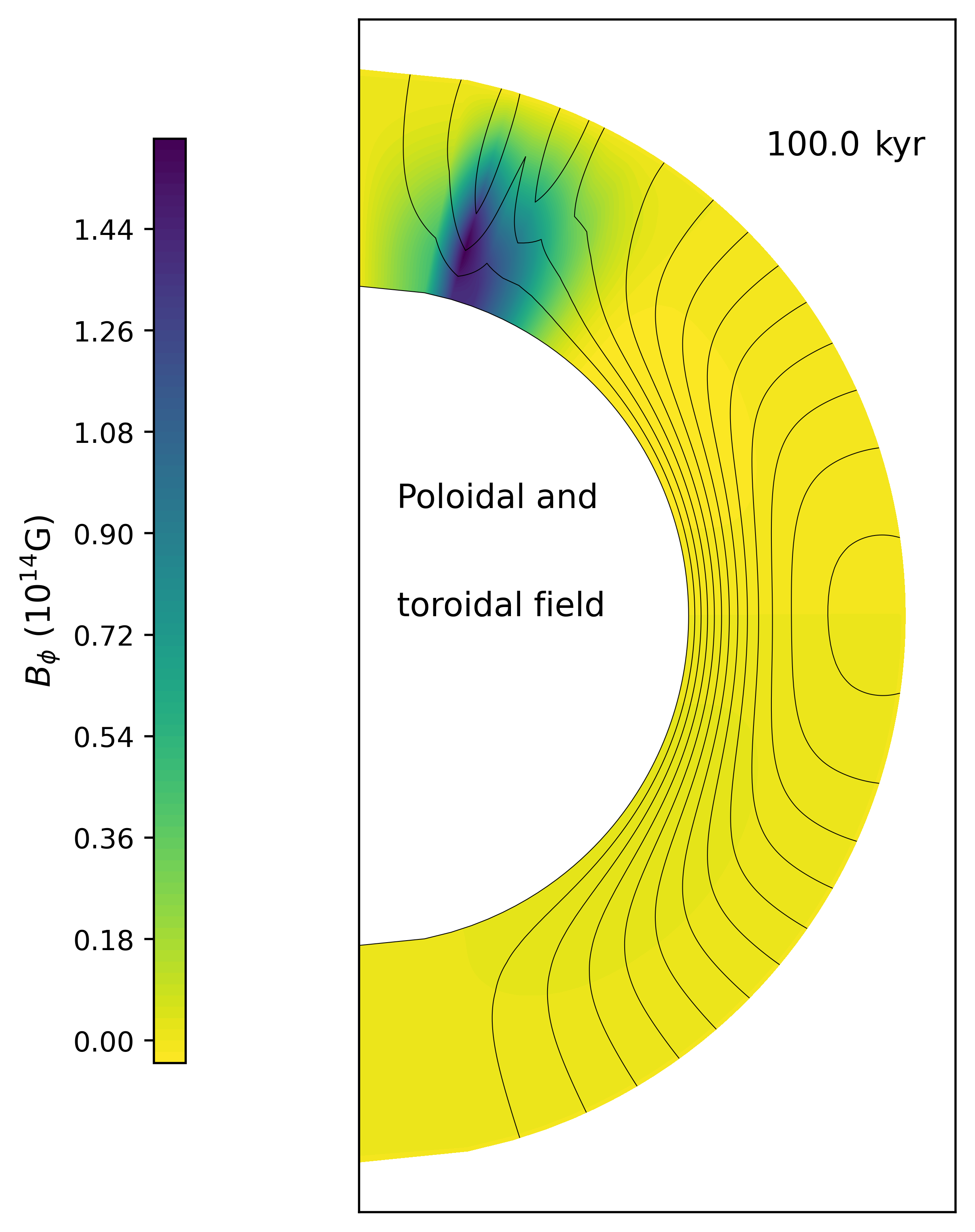}&
     c\includegraphics[width=0.3\textwidth]{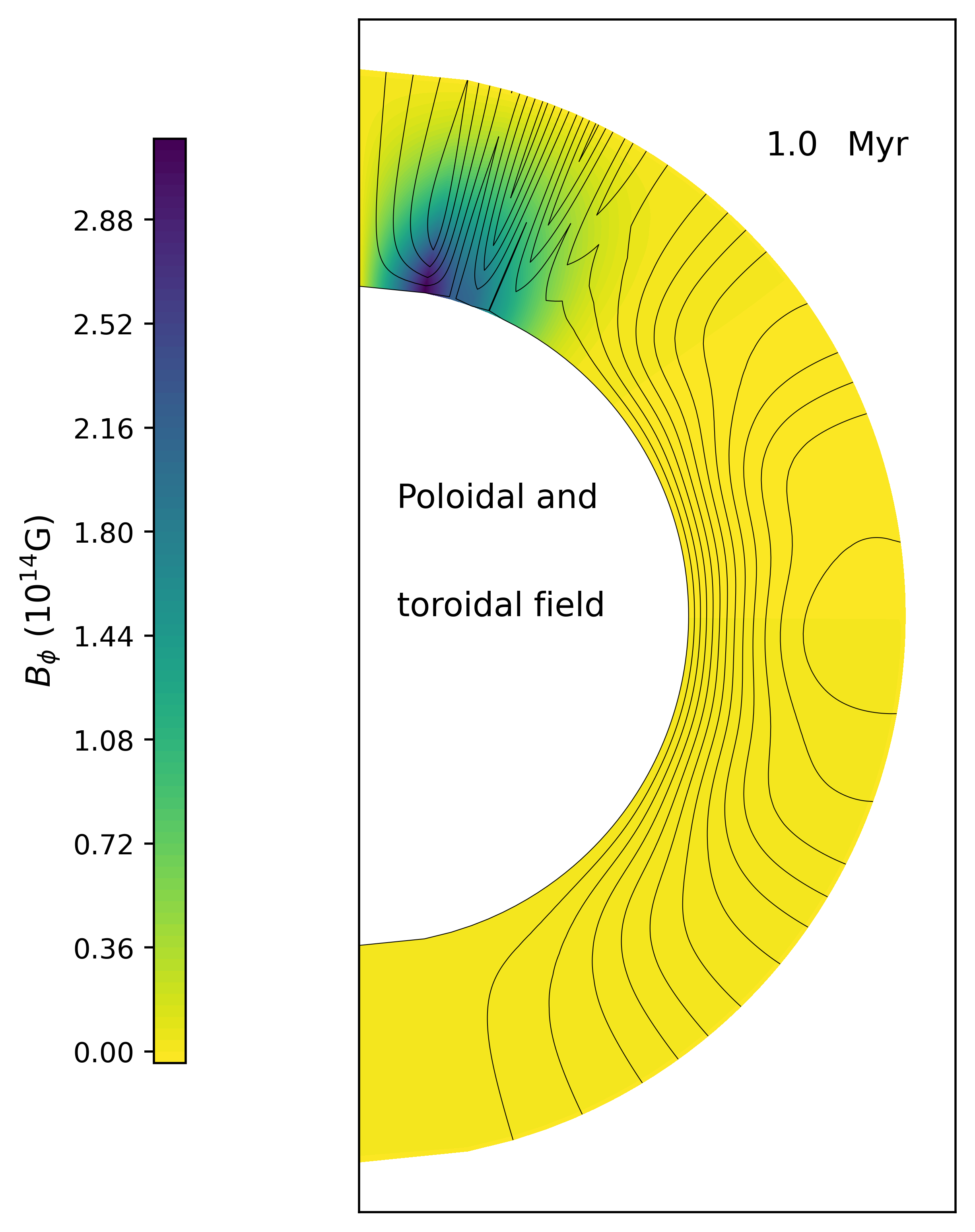}     
   \end{tabular}
\caption{Evolution of Model A through time in the case of $\Psi_1$ and $\Psi_0=10.0$. In each snapshot, the same number of poloidal lines is drawn, so that direct comparisons may be made.} 
\label{fig:2}      
\end{figure*}

\begin{figure}
\includegraphics[width=\columnwidth]{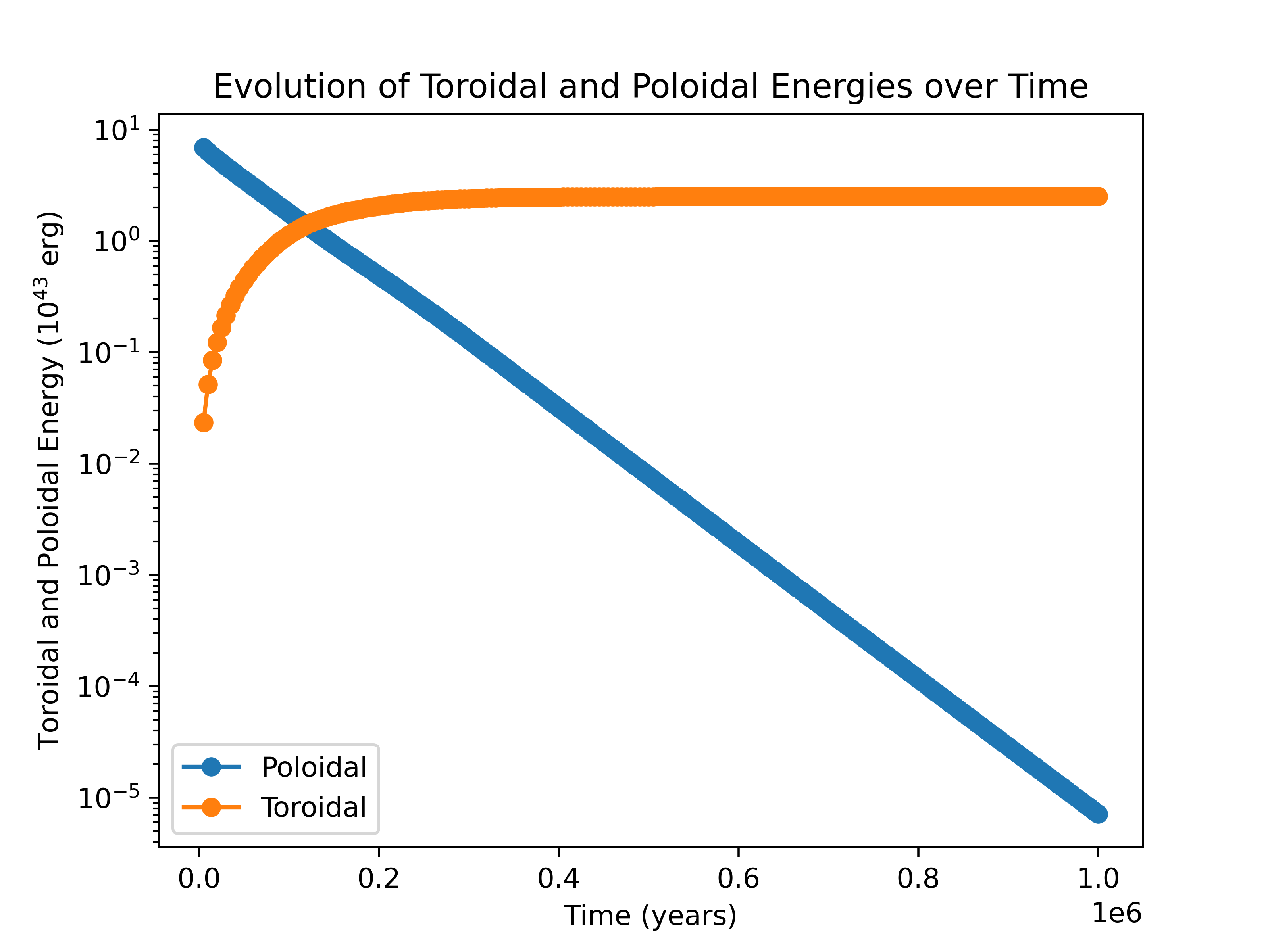}
\caption{Evolution of toroidal and poloidal energies through time of Model A in the case of $\Psi_1$, $\Psi_0=10.0$.} 
\label{fig:4}
\end{figure}

Below, we present the outputs from different models of the temperature profile. All cases presented here are simulated for a maximum time of ones million years, in order to be able to observe the saturation of the system each time.

\subsection{Model A. High temperature} \label{sec:sec4.1}
As stated in Section \ref{sec:sec3.2}, our basic layout has the following parameter values: $r_0=0.99, \mu_0=0.99$, $\sigma_r=0.1$, $\sigma_{\mu}=0.1$, and $\lambda=1$. In this subsection, we examine a rather high main crust temperature, of $T_0=10^9$ K.

We begin with an initial poloidal strength of $\Psi_0=10.0$ and the case of $\Psi_1$, eq.~(\ref{26}). Three snapshots of the evolution of the magnetic field, at 10, 100 and 1000 kyr are given in Fig.~\ref{fig:2}. The colorbar on the left refers to the value of the toroidal magnetic field, and the density of the contour lines gives an estimation of the poloidal magnetic field.

Soon after the start of the simulation, a region of strong toroidal field is formed near $(r_0,\mu_0)$, which gradually overpasses in strength the structures formed by the Hall effect (yellow and light blue regions covering the two hemispheres in Fig.~\ref{fig:2}a). As time passes, the region of strong field due to the battery shifts deeper and toward the axis of the star, as a result of velocity, resistivity, and density differences between different parts of the crust. The addition of the thermoelectric current, and its resulting toroidal field creation, cause the poloidal lines to change their initial shape, and to start to twist around the battery region. These lines eventually tend to encircle the region of the strong toroidal field, creating interesting arcades and multipoles near the pole, but are kept almost intact at distant regions.

We now turn on our attention on the evolution of the poloidal and toroidal energies through time. Their resulting values are plotted in Fig.~\ref{fig:4} as a function of time. We notice that although the poloidal energy is monotonically descreasing until reaching zero, the toroidal energy starts from zero (since no initial toroidal field is assumed) and increases until about $2.5 \times 10^{43}$ erg, at which it stabilizes.

The transition to constant values for the toroidal energies is located around 400 kyr (consistent with the analytical calculation of eq.~(\ref{23}), with the appropriate scale), at which we assume there is the saturation point. In other words, saturation occurs when the energy and field provided by the thermoelectric battery are compensated by the Ohmic losses. After saturation is achieved, the structure and values of the toroidal field remain constant, while the poloidal field decays with time.

\begin{figure*}
\begin{tabular}{p{0.5\textwidth} p{0.49\textwidth}}
  a\includegraphics[width=0.4\textwidth]{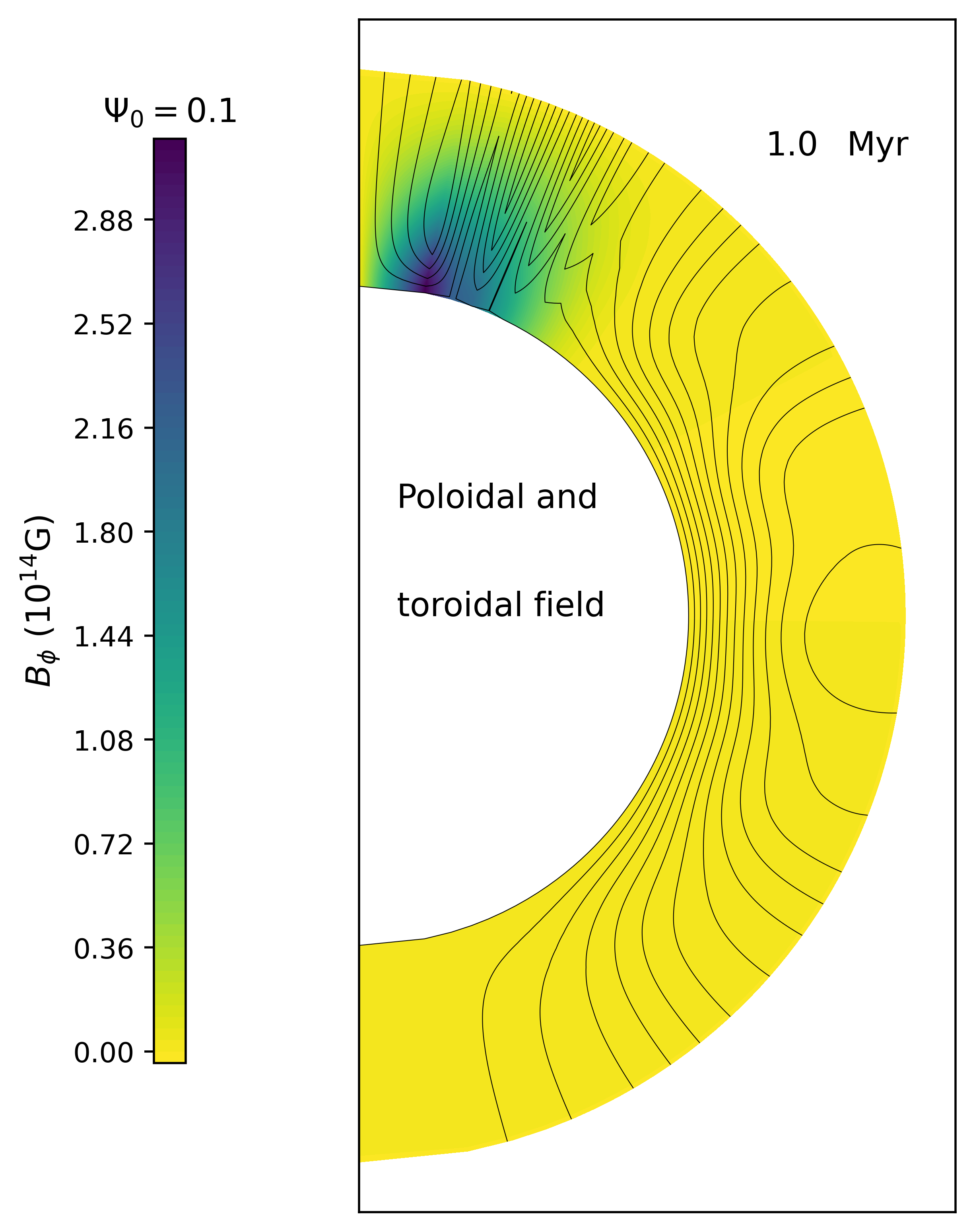}&
 b\includegraphics[width=0.4\textwidth]{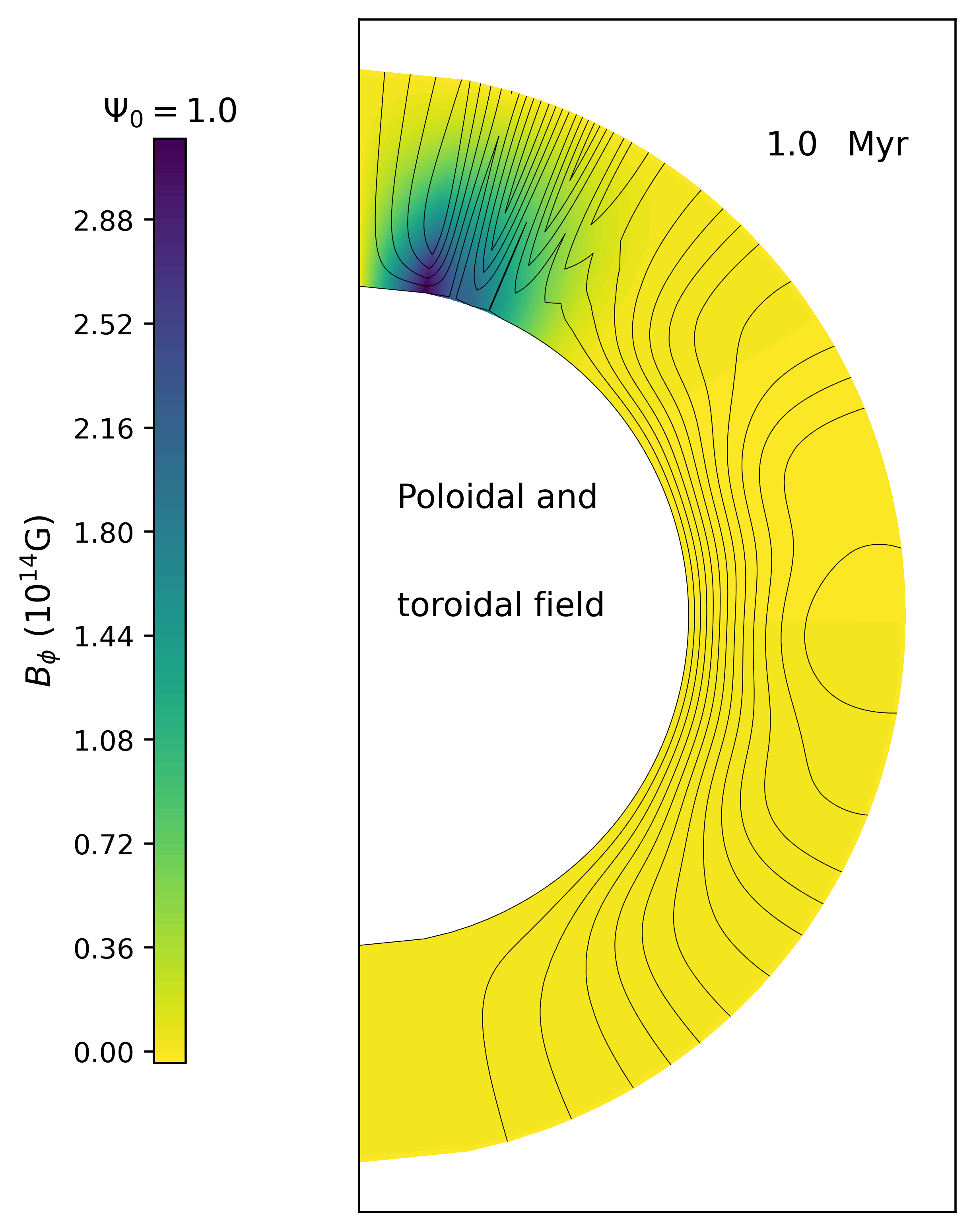}\\
  c\includegraphics[width=0.4\textwidth]{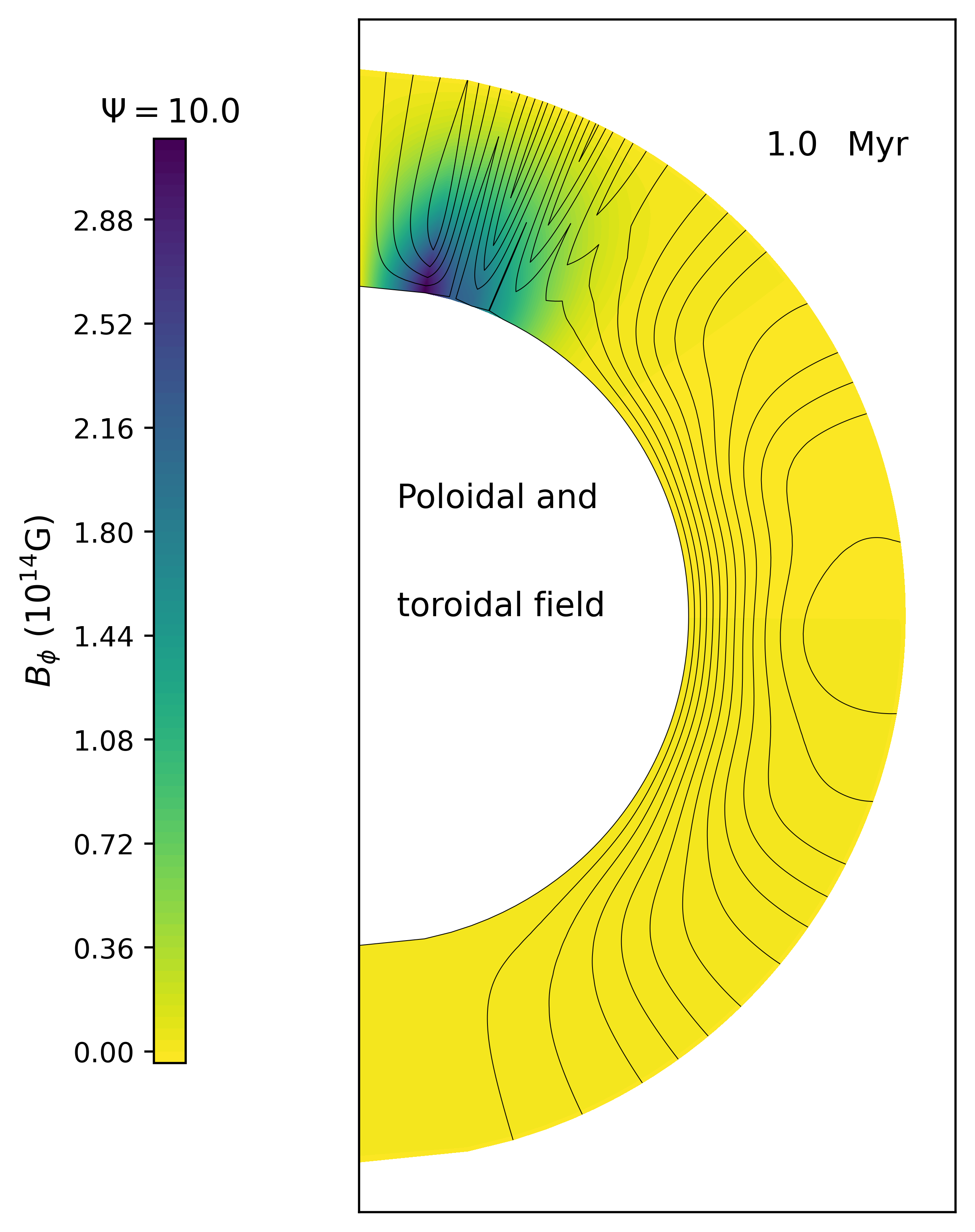}&
      d\includegraphics[width=0.4\textwidth]{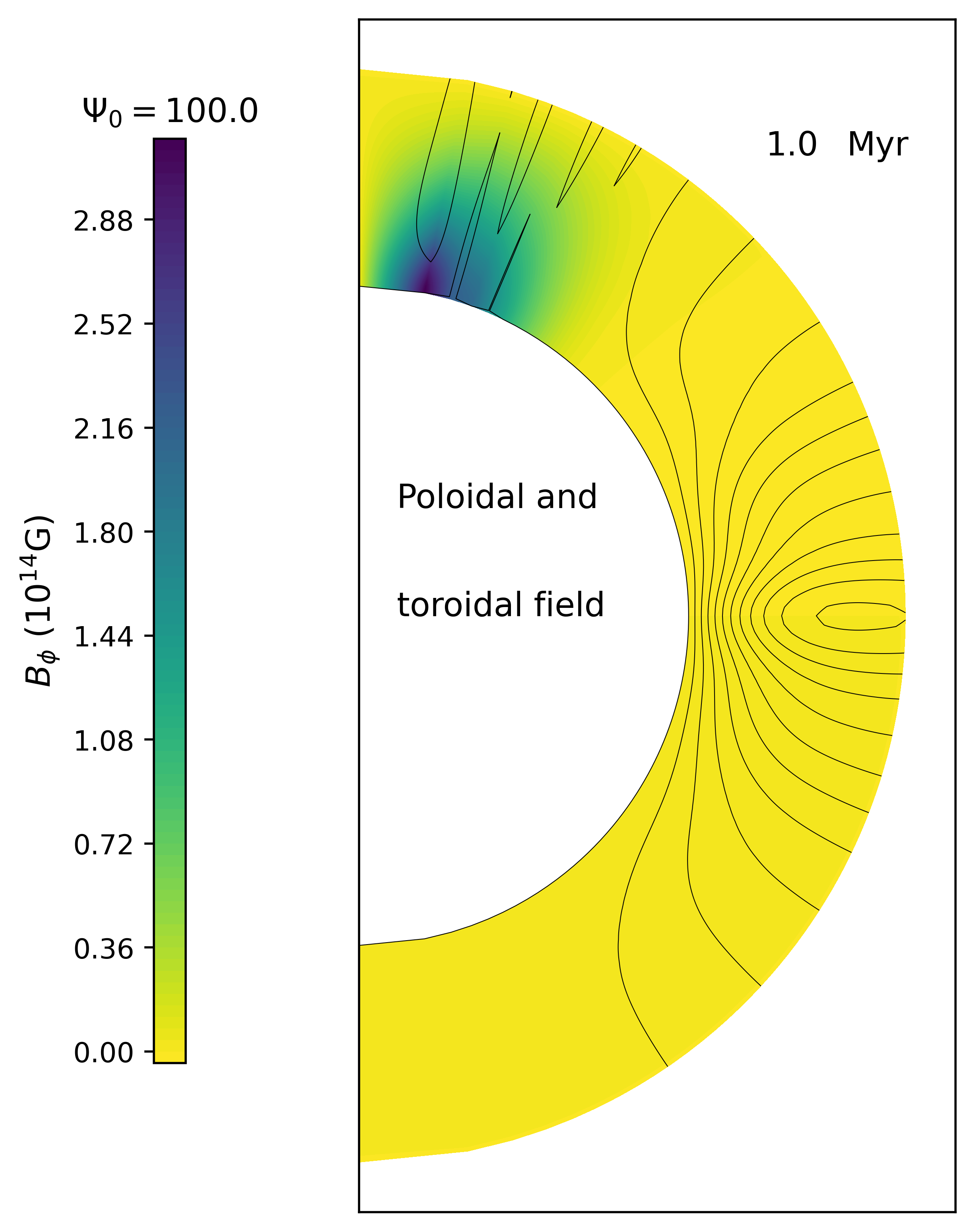} \\
     \end{tabular}
\caption{Final snapshots for Model A in the case of $\Psi_1$ for different values of $\Psi_0$ (0.1 - a, 1.0 - b, 10.0 - c, and 100.0 - d). In each snapshot, the same number of poloidal lines is drawn, so that direct comparisons may be made.}
\label{fig:3}      
\end{figure*}

We have explored the evolution of the above configuration for other values of $\Psi_0$ as well, ranging from 0.01 up to 100.0. 
The final snapshots (at simulated time of 1 Myr) are presented in Fig.~\ref{fig:3} for the cases of $\Psi_0=0.1, 1.0, 10.0$, and $100.0$. Simulations with even larger $\Psi_0$ 
were not successful. Interestingly, the larger the initial poloidal field, the sparser the final poloidal lines appear (near the battery region), but typically, the final architecture of the lines is rather similar, especially for the low-$\Psi_0$ cases. Additionally, all poloidal energies gradually drop to zero, despite starting at different values depending on $\Psi_0$. 

Regarding the toroidal field, it appears that the battery term dominates immediately over the Hall term for lower $\Psi_0$, whereas it takes more time to achieve that for higher values. Still, we note that saturation is observed at roughly the same time for all of them, while the final value of the maximum toroidal field in the star is also the same, $3.16 \times 10^{14}$ G, as shown in Fig.~\ref{fig:3}. This value is representative for magnetars, while the remaining star is described by values around $-5 \times 10^{12}$ G for $\bm{B}_{T}$, which is two orders of magnitude lower. This small negative toroidal field is a result of the Hall term, which causes oscillations of the poloidal lines and diverts some field along the $\phi$-axis; the battery term does not contribute far away from the main thermal gradient area. In addition, for all choices of $\Psi_0$, the toroidal energy also reaches the same steady value, but for $\Psi_0 \gtrsim 10$
, the toroidal energy reaches a peak before dropping at this value. 

We have also experimented with even higher temperatures. For instance, setting $T_0=5 \times 10^9$ K resulted in a maximum of $\bm{B}_{T}=2.35 \times 10^{15}$ G, noticing even sparser structure of the poloidal lines. 
However, since even higher temperatures are not physically acceptable for the crusts of even highly magnetized magnetars, we do not consider this scenario further. 

\begin{figure*}
\begin{tabular}{p{0.3\textwidth} p{0.3\textwidth} p{0.3\textwidth}}
  a\includegraphics[width=0.3\textwidth]{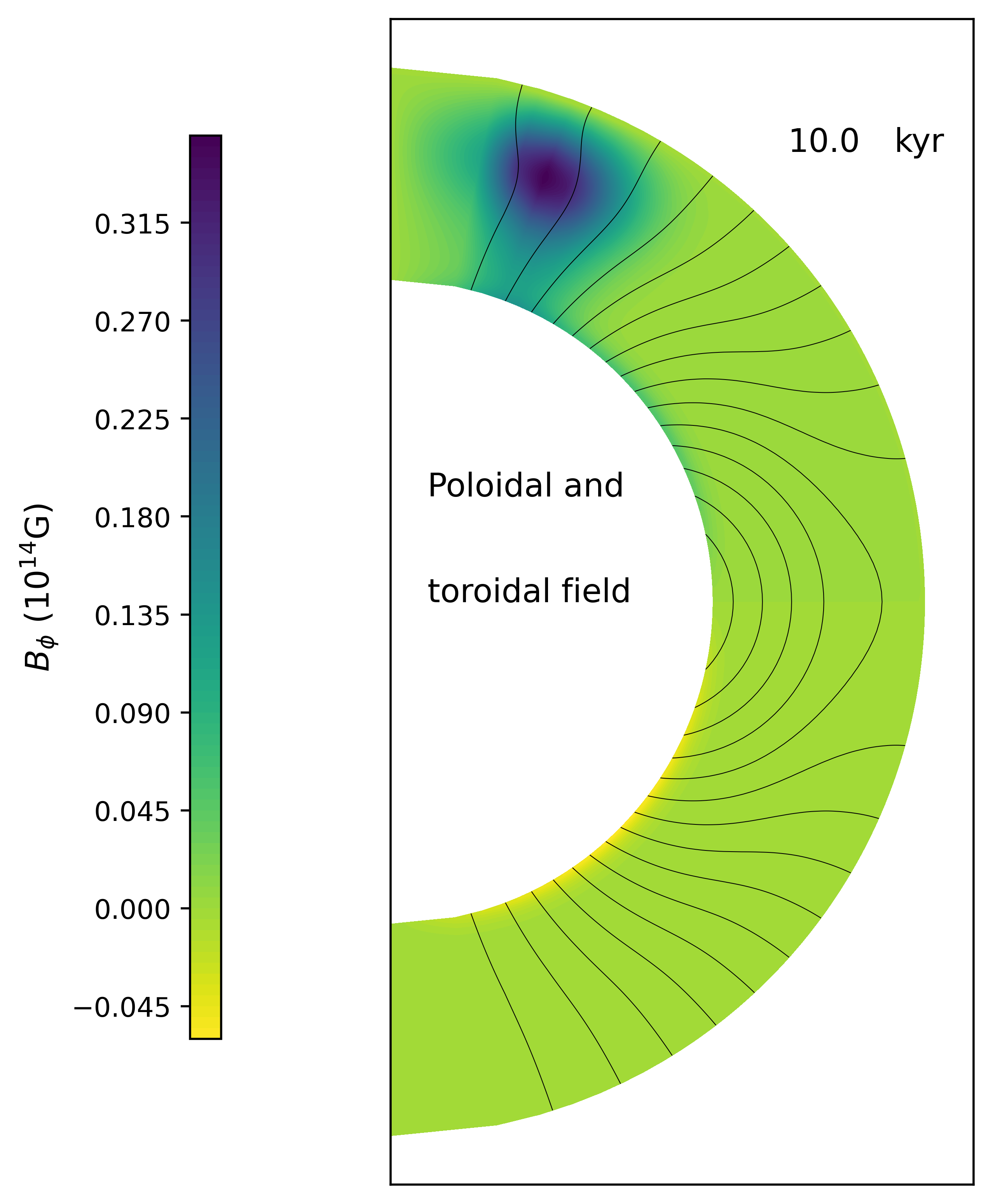}&
    b\includegraphics[width=0.283\textwidth]{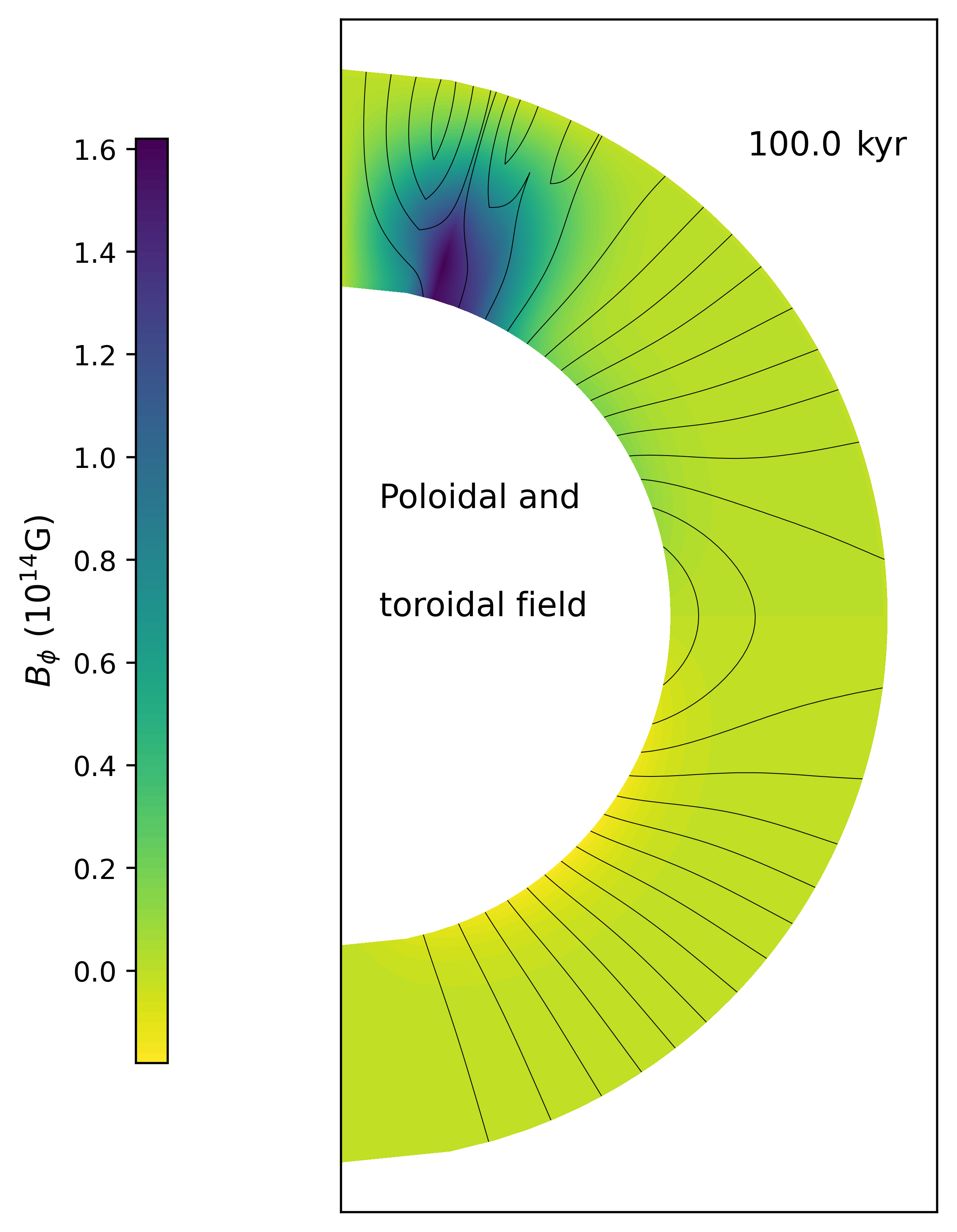}&
     c\includegraphics[width=0.29\textwidth]{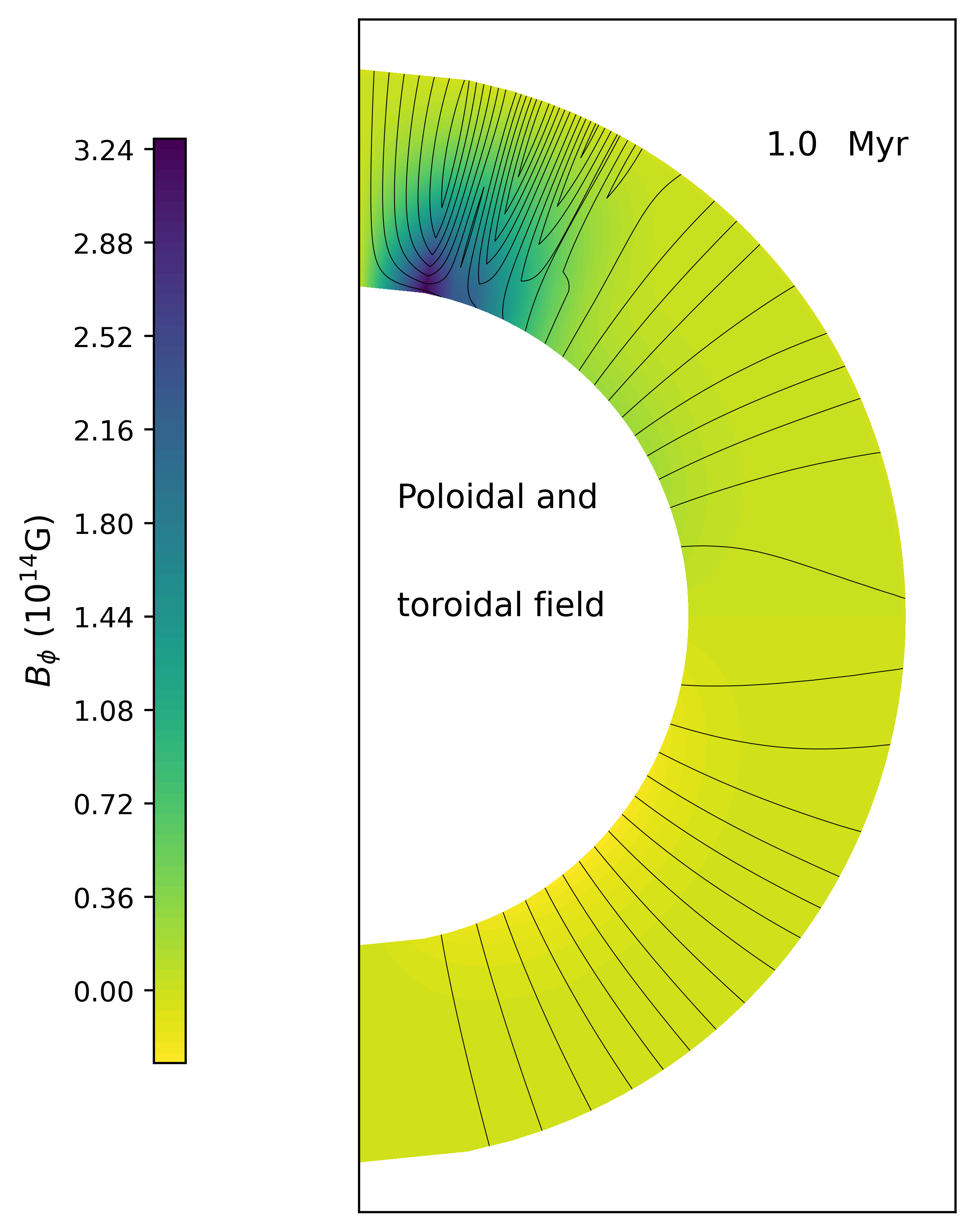} 
     \end{tabular}
\caption{Evolution of Model A through time in the case of $\Psi_2$ and $\Psi_0=0.01$. In each snapshot, the same number of poloidal lines is drawn, so that direct comparisons may be made.}
\label{fig:5}      
\end{figure*}

\begin{figure}
\includegraphics[width=\columnwidth]{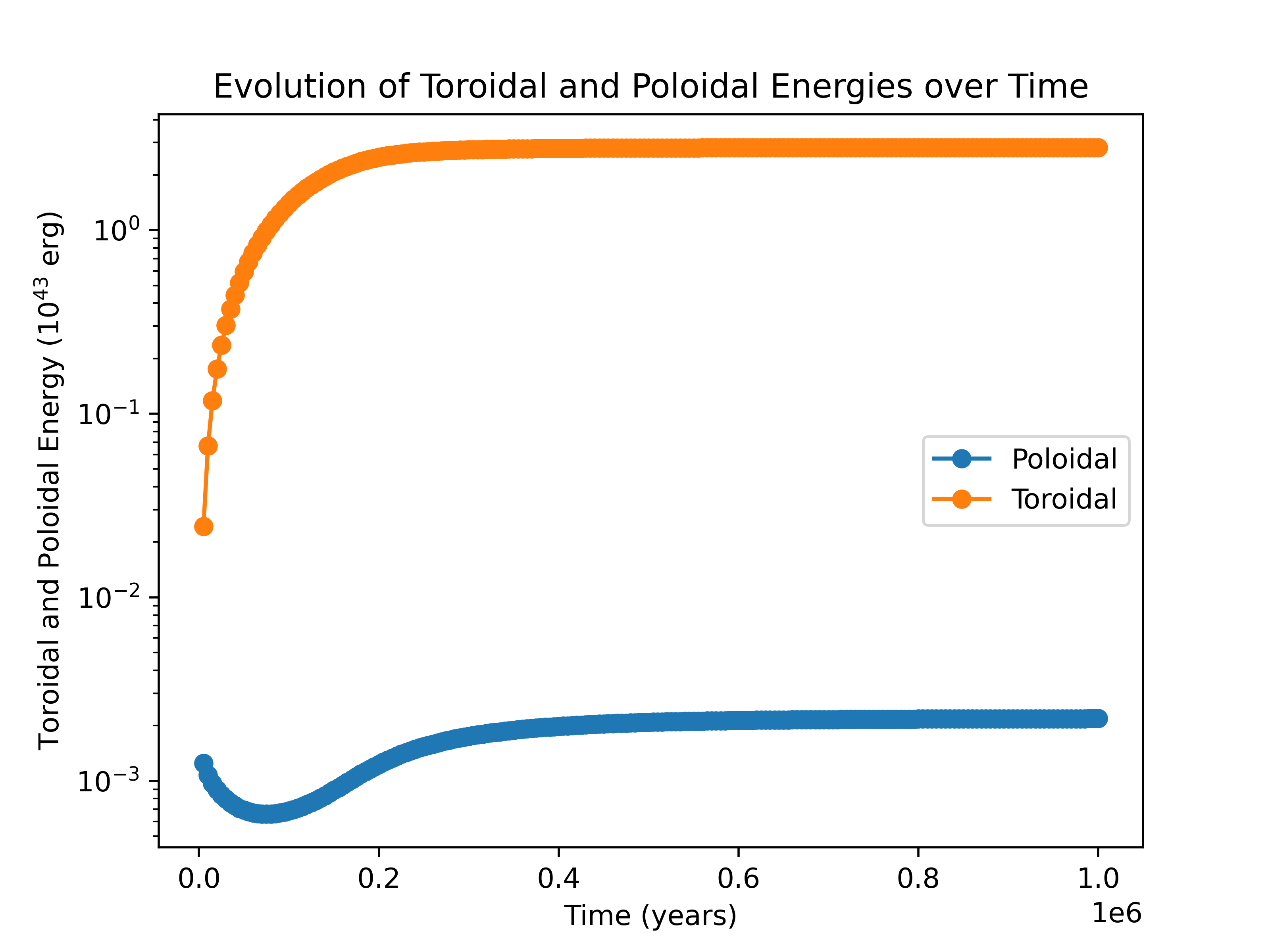}
\caption{Evolution of toroidal and poloidal energies through time of Model A in the case of $\Psi_2$, $\Psi_0=0.01$.}
\label{fig:7}
\end{figure}

The case of $\Psi_2$, eq.~(\ref{27}) is different. The simulations are only successful for very low initial values of $\Psi_0$, as in the opposite case features of an extremely large field are created, rendering the code unstable. Hence, we simulate temperatures of $T_0=10^9$ K and $\Psi_0=0.01$ (given in Fig.~\ref{fig:5}), or even lower. The respective evolution of the energies is shown in Fig.~\ref{fig:7}. The high-$\bm{B}_{T}$ region is observed to shift to inner depths and smaller polar angles. At roughly the same time as in the previous configurations, saturation is again achieved (though at a slightly higher toroidal energy value, while the poloidal energy also maintains a constant value after $\sim$300 kyr because the poloidal flux is non-zero at the crust-core interface), apparently stabilizing the geometry of poloidal lines after $\sim$700 kyr. The maximum value of the toroidal field is now slighlty higher ($3.27 \times 10^{14}$ G) and the minimum ones around $-2.50 \times 10^{13}$ G. 

\subsection{Model B. Low temperature} \label{sec:sec4.2}
\begin{figure*}
\begin{tabular}{p{0.5\textwidth} p{0.49\textwidth}}
 a\includegraphics[width=0.4\textwidth]{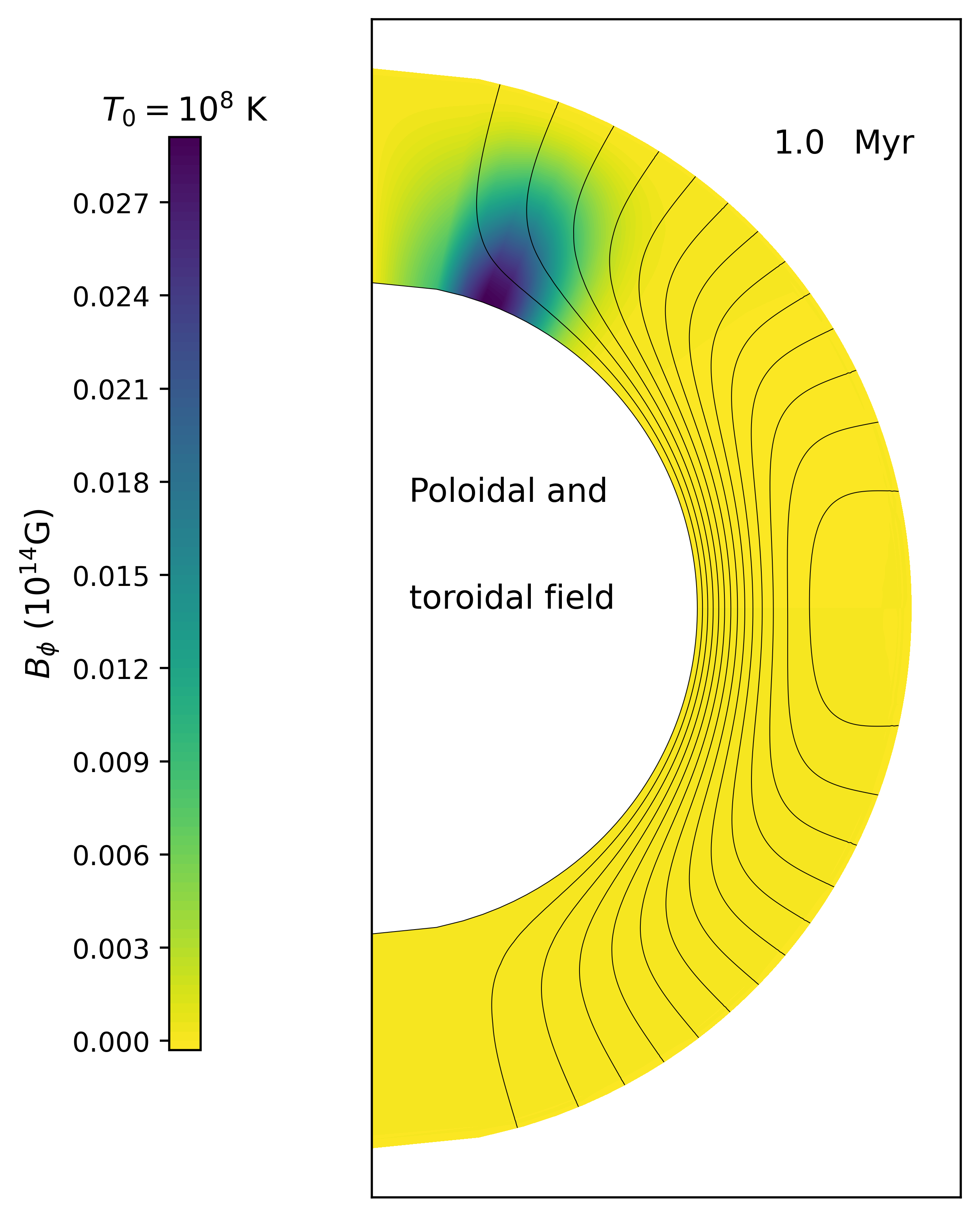}&
    b\includegraphics[width=0.4\textwidth]{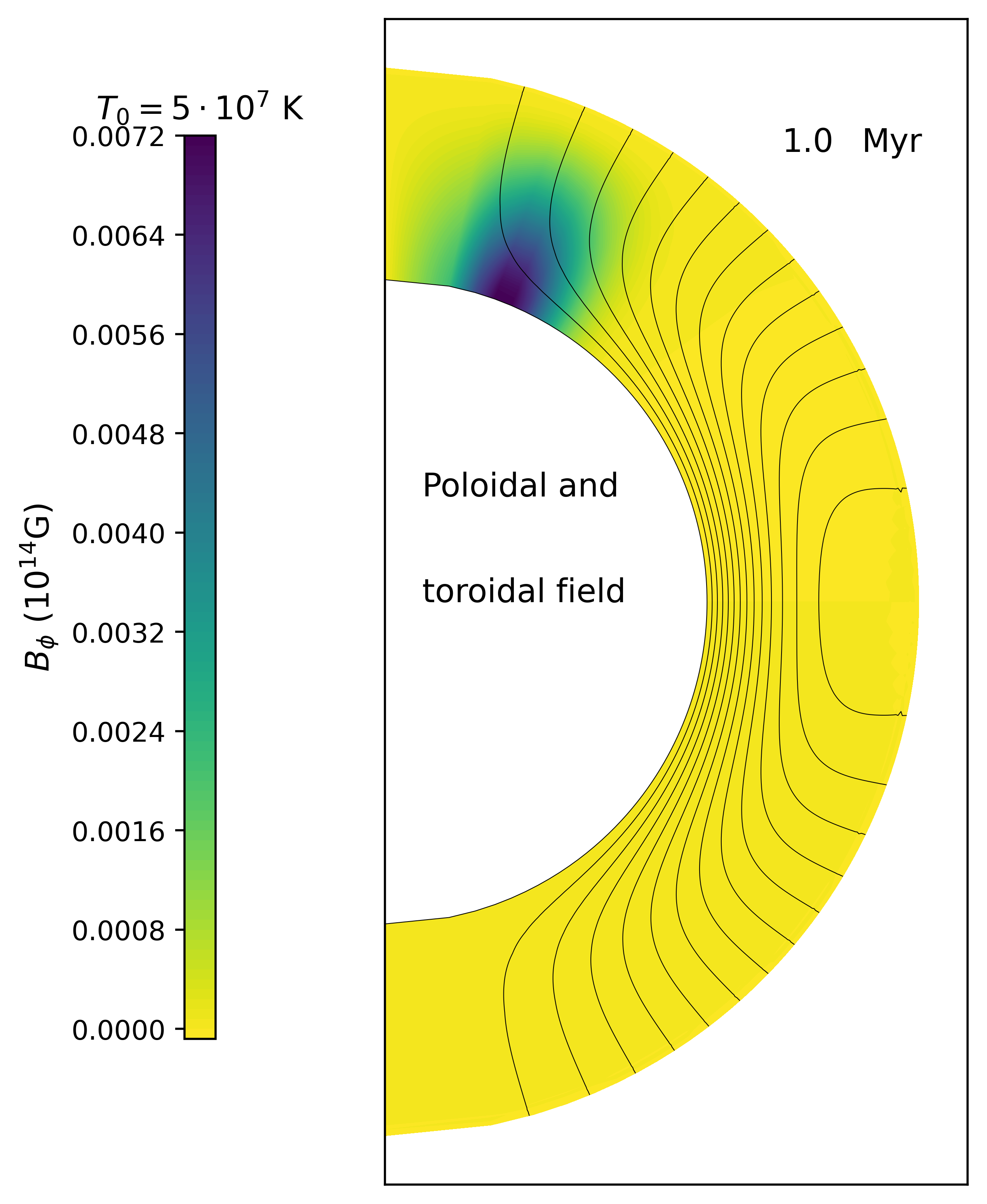}\\
     \end{tabular}
\caption{Final snapshots for Model B in the case of $\Psi_1$ and $\Psi_0=1.0$ for $T_0=10^8$ K (a) and $T_0=5 \cdot 10^7$ K (b).}
\label{fig:11}      
\end{figure*}

Keeping the same set of parameters as in Section \ref{sec:sec4.1} ($r_0=0.99, \mu_0=0.99$, $\sigma_r=0.1$, $\sigma_{\mu}=0.1$, $\lambda=1$), we now consider lower crust temperatures, namely $T_0$ being between $10^7$ and $10^8$ K. Starting with the case of $\Psi_1$, Fig.~\ref{fig:11}a presents the final snapshot (1 Myr) when $T_0 = 10^8$ K and $\Psi_0=1.0$. Having examined initial poloidal fields $\Psi_0$ in the range 0.01 to 10.0 (for lower temperatures the battery term in the resulting toroidal field can only prevail over weak poloidal fields
, the final outputs are insensitive to this choice, all reaching a steady state of $2.92 \times 10^{12}$ G at most for the toroidal field. At this low value, the toroidal field is not strong enough to twist the existing poloidal lines, hence no such feature is observed now, apart from minor oscillations after about 700 kyr. As a result, the toroidal energy rises to values of $\sim$$10^{39}$ erg, while the poloidal energy follows a similar behavior as in Model A, diminishing to zero as time passes. The remaining star (far from the battery region) is maintained at fields $\sim$$-10^{10}$ G, caused by the interplay between Hall drift and Ohmic dissipation alone. 

As expected from eq.~(\ref{1}), lower temperatures lead to even lower thermally induced magnetic fields. We simulated a temperature $T_0 = 5 \times 10^7$ K, and the final result in the case of $\Psi_1$ and $\Psi_0=1.0$ 
is shown Fig.~\ref{fig:11}b. $\bm{B}_{T}$ has a value of $7.2 \times 10^{11}$ G near the pole, which is weaker by almost an order of magnitude than the field in the case of Fig.~\ref{fig:11}a, as is the toroidal energy after saturation.

We notice that for the temperatures examined here ($10^7$ - $10^8$ K), the toroidal fields are in the range of $\sim$$10^{11}$ – $10^{12}$ G, which is representative for typical pulsars. At even lower temperatures, the mechanism becomes rather ineffective. In particular, for $T_0 \lesssim 10^7$ K no battery feature is apparent and the resulting magnetic field evolution is mainly described by the combination of Hall and Ohm terms.

In the field $\Psi_2$ scenario, the Hall term outweighs battery even at $10^8$ K because no signs of the latter are observed. We considered $\Psi_0$ as low as 0.01, since as explained in Section \ref{sec:sec4.1}, the code cannot sustain higher values. The case of radial field assumed to penetrate the core is therefore not considered further.

\subsection{Experiments with the temperature function} \label{sec:sec4.3}
This subsection is dedicated to modifying the parameters kept constant ($r_0$, $\mu_0$, $\sigma_r$, $\sigma_{\mu}$, $\lambda$) and to exploring their effect on the results. Unless otherwise stated, we will adopt as our main simulation setup the one presented in Figs.~\ref{fig:2}, ~\ref{fig:4} (Model A, $\Psi_1$, $\Psi_0=10.0$, $T_0 = 10^9$ K) and will make any changes (one at a time) regarding the temperature function parameters to this one.

\begin{figure}
\includegraphics[width=0.44\textwidth]{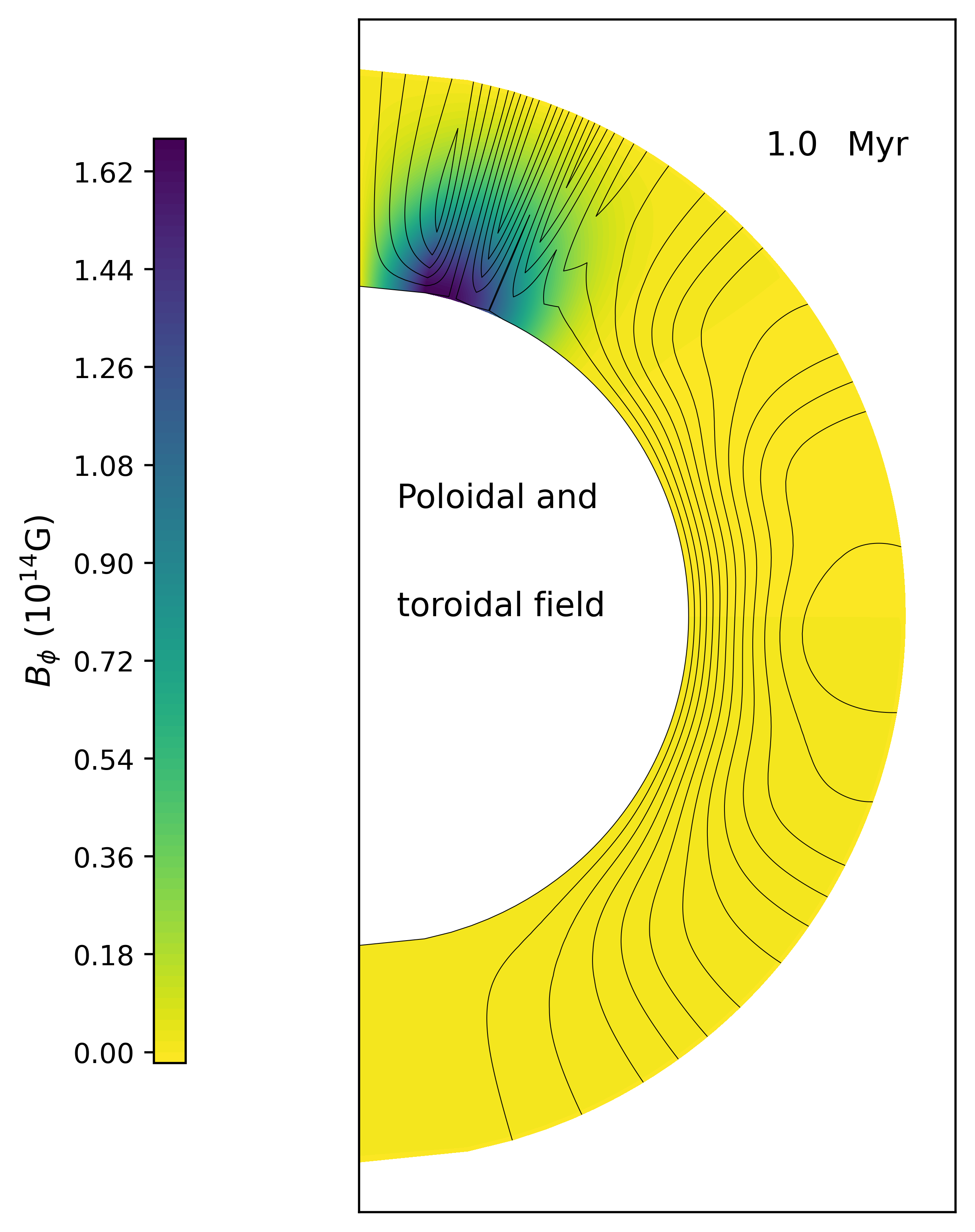}
\caption{Final snapshot for Model A in the case of $\Psi_1$ and $\Psi_0=10.0$ for $r_0=0.90$.}
\label{fig:9}
\end{figure}

We first experiment with the position of the point with the highest temperature throughout the crust (referred to as the heat point). We assumed so far that the maximization of temperature occurs in $r_0 = 0.99$. The reason behind this option lies in the fact that magnetospheric currents (which contribute to the augmentation of temperature near the polar cap) are attenuated within the first $\sim$100 m of the star (at a depth corresponding to approximately 1\% of the stellar radius). While this seems a realistic choice, heat might accumulate at other points inside of this. Consequently, we ran two tests with $r_0 = 0.95$ and $r_0 = 0.90$. In the former case, the results appear to be almost identical compared to when $r_0 = 0.99$ in terms of the final values and the architecture of the toroidal and poloidal field, their energies, and saturation times. Setting $r_0 = 0.90$ (the heat point is located at the base of the crust) some changes occur, however. The final snapshot, at 1 Myr, is given in Fig.~\ref{fig:9}, in which the same number of poloidal lines as in Fig.~\ref{fig:2} are drawn for comparison. The highest $\bm{B}_{T}$ values appear to have dropped to almost half ($\sim$$1.68 \times 10^{14}$ G), as does the toroidal energy (about $1.0 \times 10^{43}$ erg). By examining the time evolution, we additionally conclude that saturation occurs approximately 100 kyr later.

\begin{figure}
\includegraphics[width=0.45\textwidth]{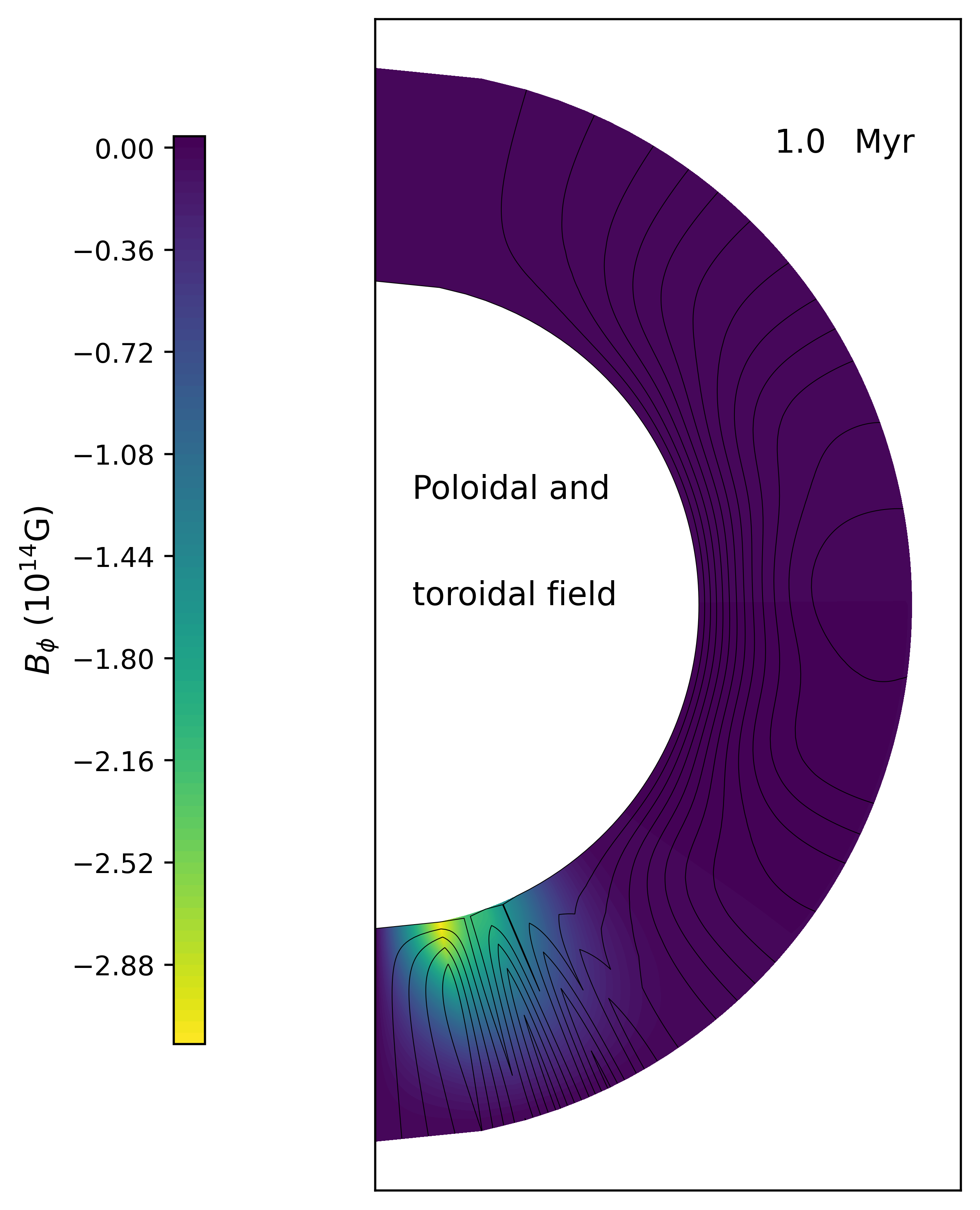}
\caption{Final snapshot for Model A in the case of $\Psi_1$ and $\Psi_0=10.0$ for $\mu_0=-0.99$.} 
\label{fig:8}
\end{figure}

We simulate the symmetric configuration for the position of the heat point in the meridional (symmetric with respect to the equator), that is, $\mu_0=-0.99$. Fig.~\ref{fig:8} shows the final output (again the same number of poloidal lines as in Fig.~\ref{fig:2} are drawn). Clearly, the heat point located at the south pole yields the opposite values (i.e., the battery region is surrounded by strong negative magnetic fields), but the absolute values remain the same. Moreover, the same occurs with the other results (saturation time and magnetic energies). Despite that, moving the heat point to other latitudes (e.g., to the equator or to mid-latitudes) was not possible. The development of a strong toroidal field ($10^{14}$ G) in regions far from the poles twists the existing poloidal lines in narrow areas very strongly, which prevents the code from working properly. 

\begin{figure*}
\begin{tabular}{p{0.5\textwidth} p{0.49\textwidth}}
 a\includegraphics[width=0.4\textwidth]{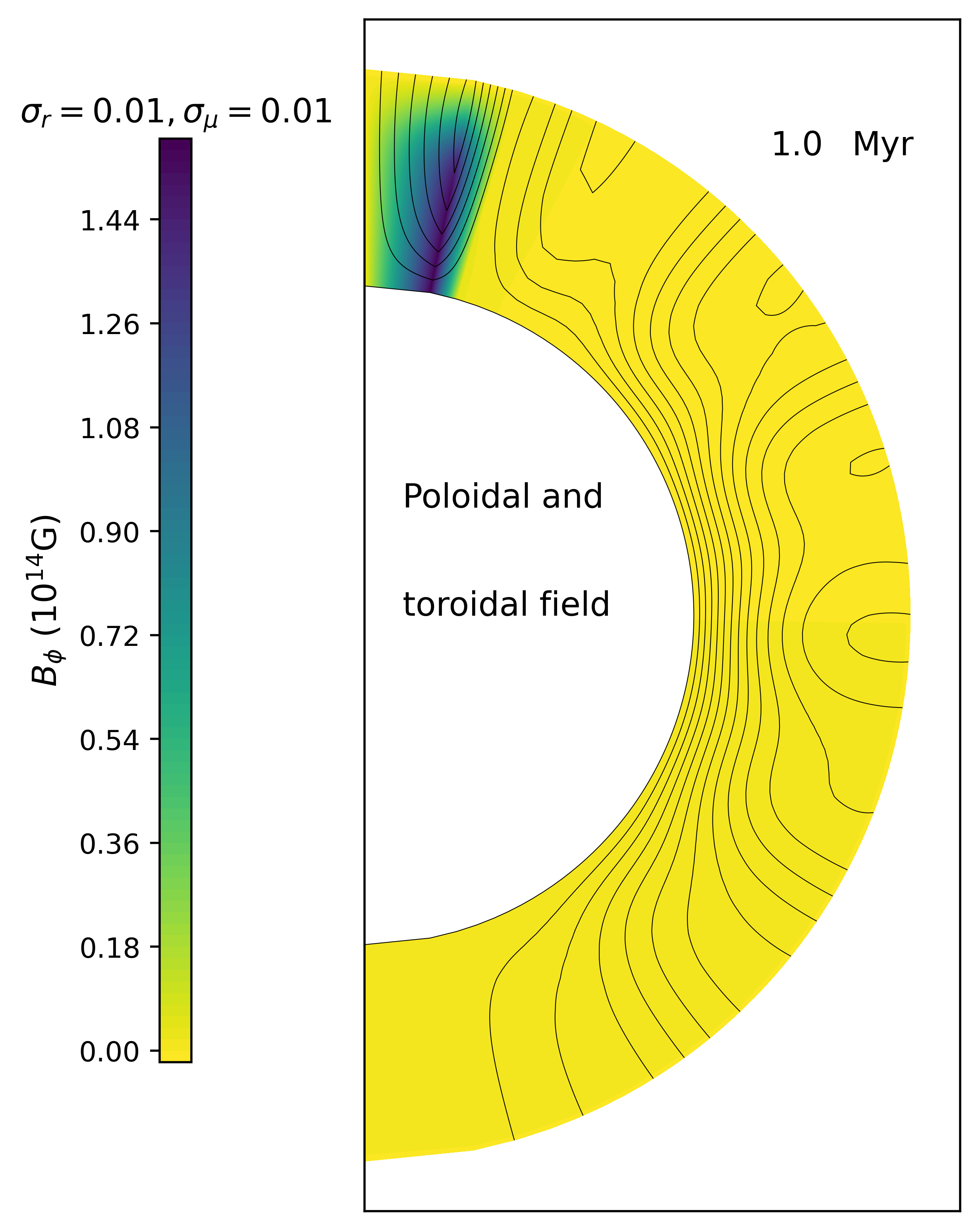}&
    b\includegraphics[width=0.4\textwidth]{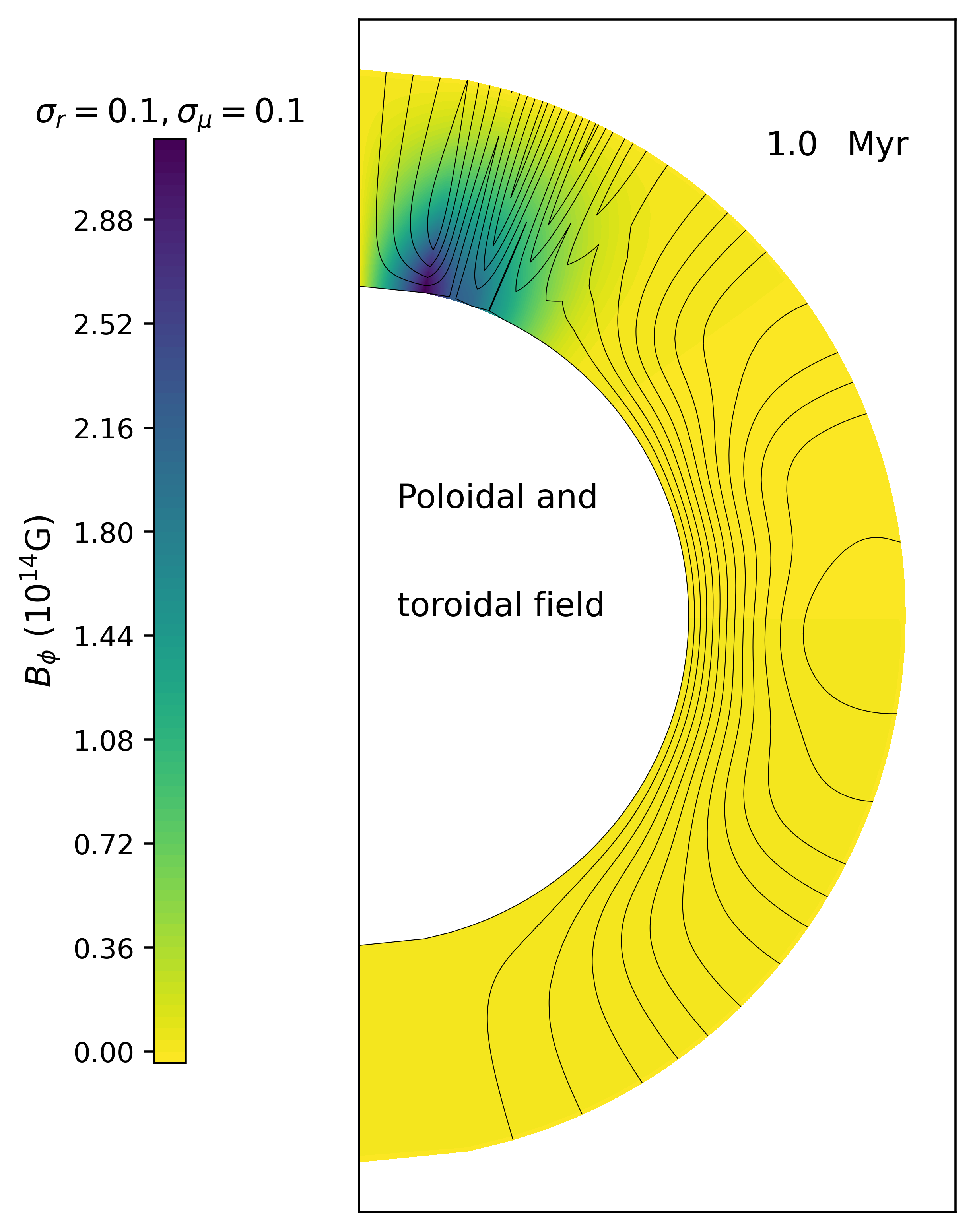}\\
     c\includegraphics[width=0.4\textwidth]{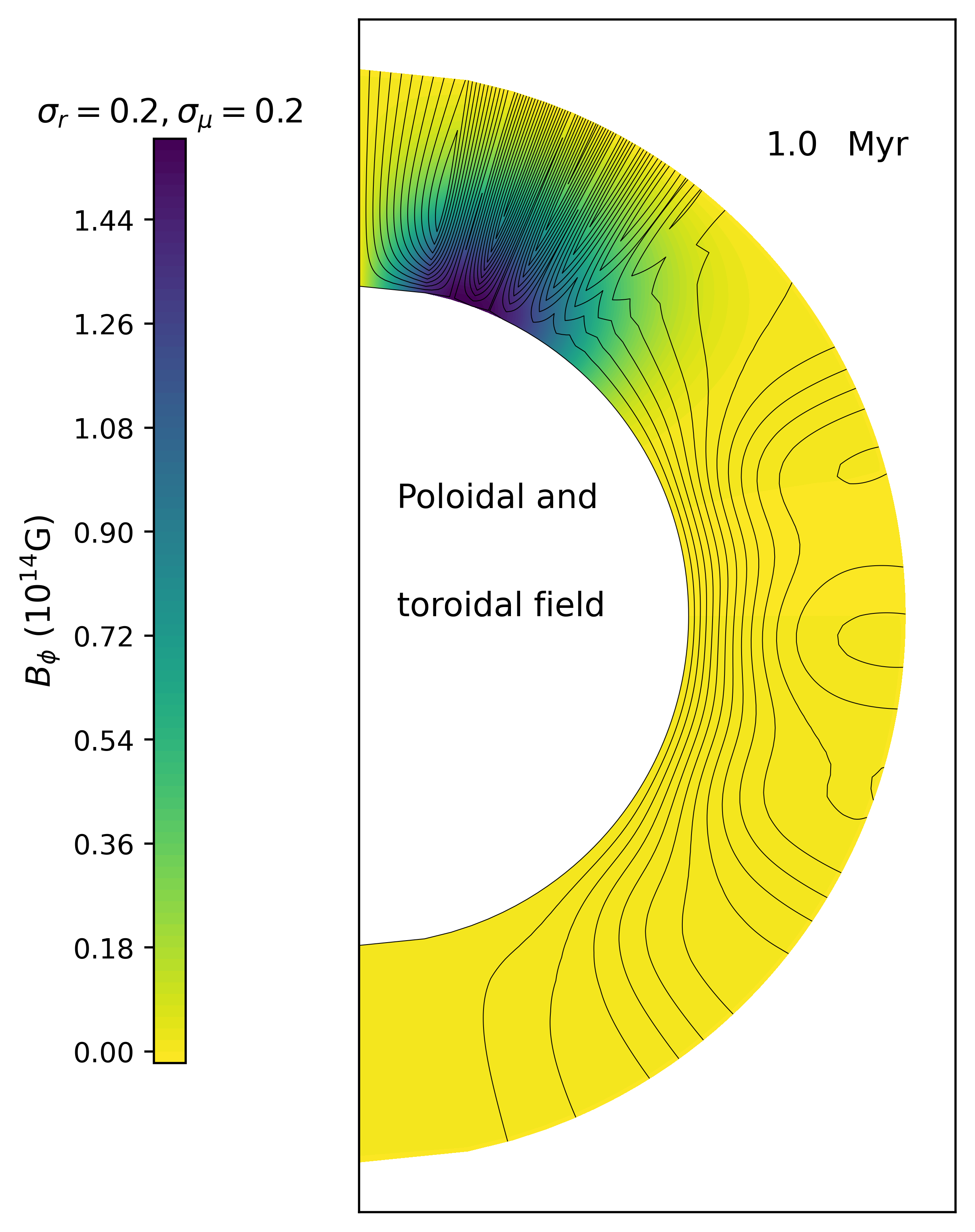} &
    d\includegraphics[width=0.4\textwidth]{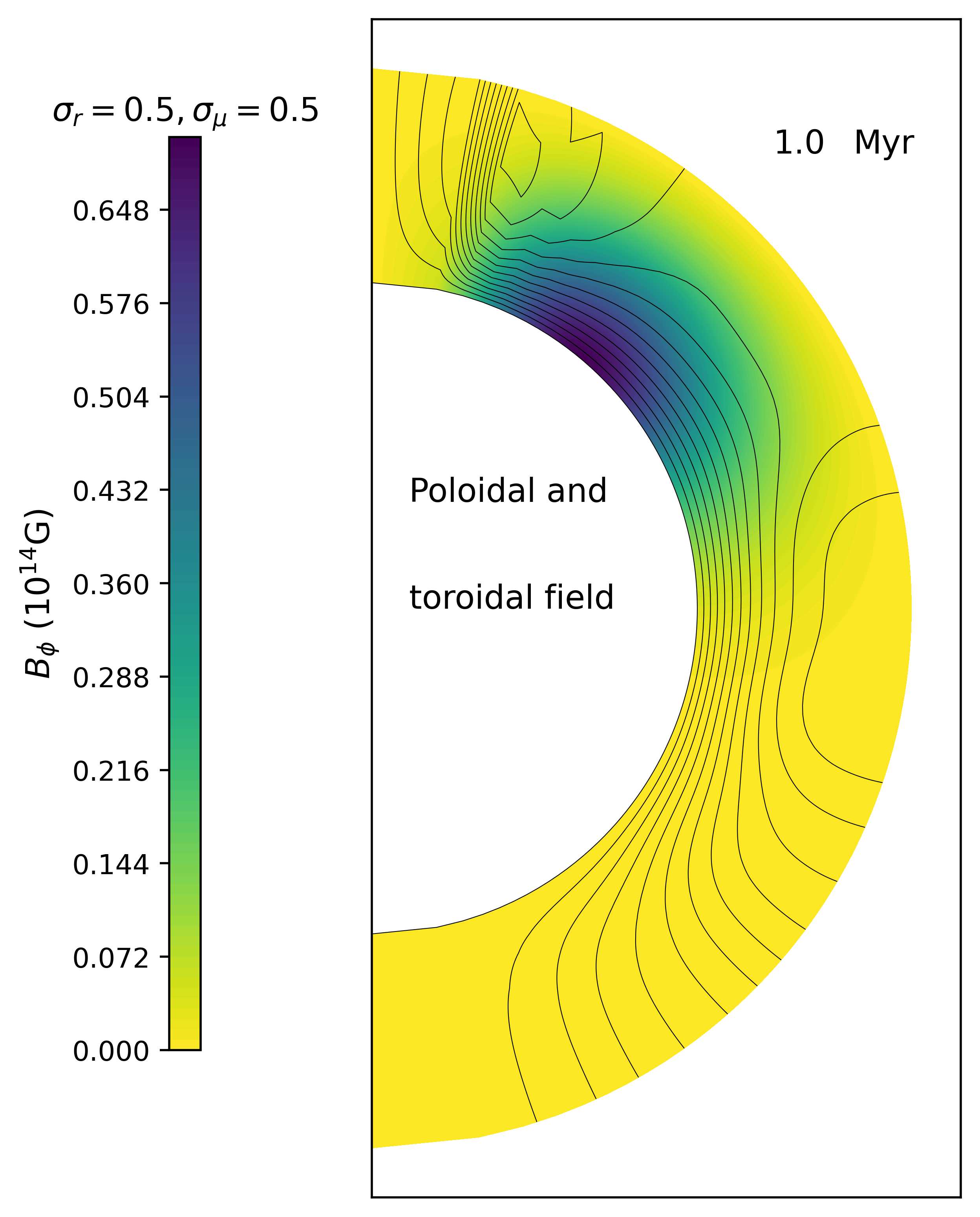} \\
     \end{tabular}
\caption{Final snapshots for Model A in the case of $\Psi_1$ and $\Psi_0 = 10.0$ for 4 different values of $\sigma_r$ and $\sigma_{\mu}$: (0.01 - a, 0.1 - b, 0.2 - c, and 0.5 - d).} 
\label{fig:6}      
\end{figure*}

Next, we test the spread of the Gaussian function in $r$ and $\mu$. 
By simulating an idealized neutron star, we have the freedom to vary the values $\sigma_r$ and $\sigma_{\mu}$, as the actual temperature profile of a neutron star is unknown. We choose to examine three additional scenarios. The first one is the most conservative, at which $\sigma_r$ and $\sigma_{\mu}$ are reduced ten times, at 0.01. We then double these parameters, at 0.2, while an even more optimistic scenario, in which these parameters are as large as 0.5 each, is examined as well.

These three cases, along with the default case, are presented in Figs.~\ref{fig:6}a-d at a simulated time of 1 Myr (all of the figures have the same number of poloidal lines). In principle, the small yellow region of strong toroidal field is shifted, as expected, because different spreads also change the point of the highest temperature gradient. Low standard deviations only restrict the battery region near the pole, but with multipoles evident farther from it. The highest $\bm{B}_{T}$ values look similar for $\sigma_r$, $\sigma_{\mu}$ equal to 0.01 and 0.2. The evolution in the latter case reveals that around 400 kyr, the poloidal lines start to increase in density forming wavy structures, whereas the toroidal field reaches a maximum value ($1.83 \times 10^{14}$ G) that has more than one region with a strong field. These fields are later unified as the field drops and saturates at values $1.58 \times 10^{14}$ G at roughly 600 kyr. The large spread of the last case, Fig.~\ref{fig:6}d, along with the fact that the point of the highest temperature gradient now lies in the core, have as a consequence that the toroidal field is even lower, $7.0 \times 10^{13}$ G, and more time (about 900 kyr) is required for saturation. Lastly, we notice that for all of these configurations, field lines become almost radial and parallel around it, because of twisting in narrow region near the initial heat point.

\begin{figure}
\includegraphics[width=0.45\textwidth]{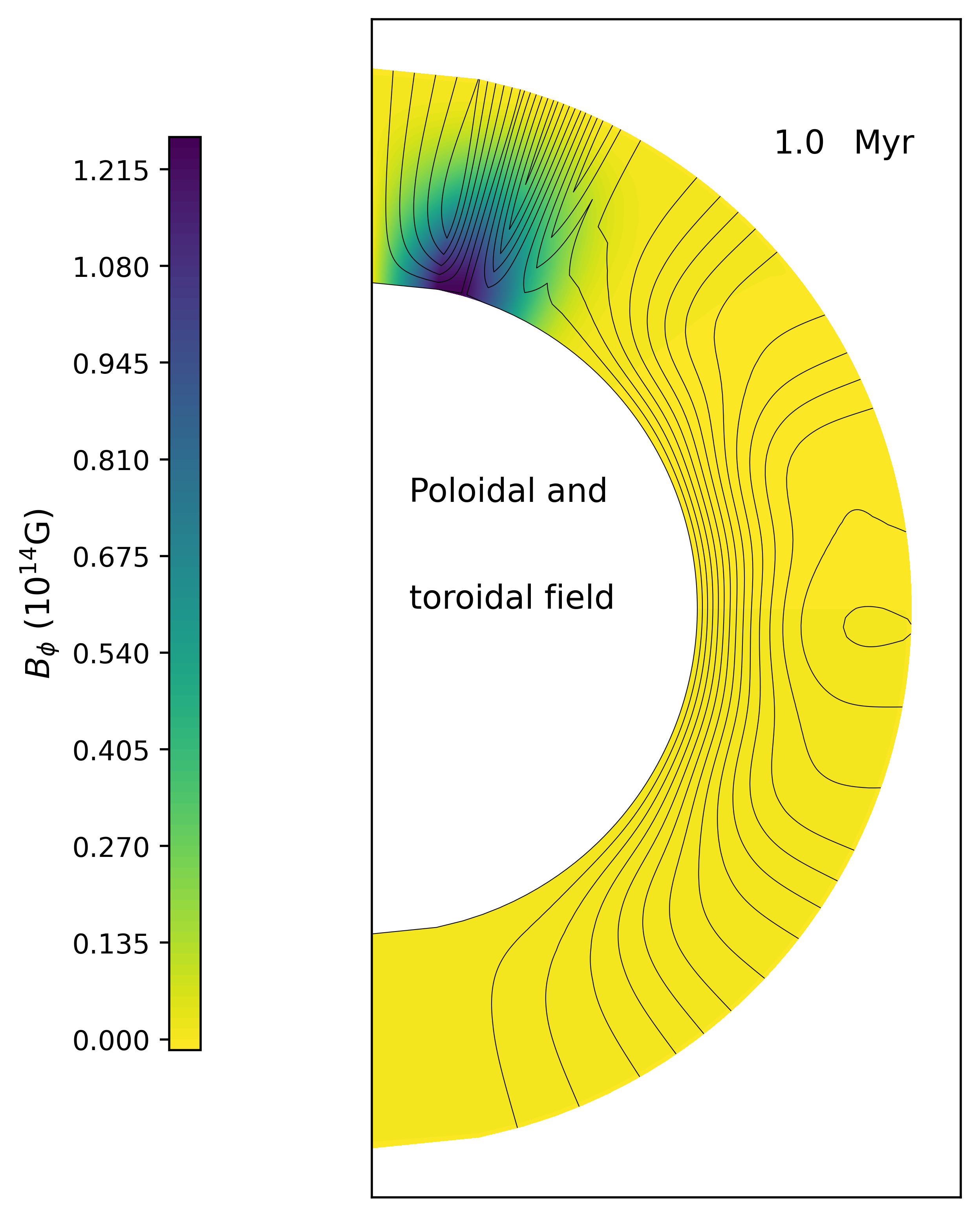}
\caption{Final snapshot for Model B in the case of $\Psi_1$, $\Psi_0=10.0$ and $T_0=10^8$ K for $\lambda=10$.}
\label{fig:10}
\end{figure}

Moving on, we briefly discuss the parameter $\lambda$. We consider the following situation: the typical temperature is set at $T_0=10^8$ K (Model B), which is a more realistic temperature for a neutron star, but we assume that the temperature rose 11 times at the heat point, that is $\lambda=10$. The final figure, with the remaining model choices being $\Psi_1$ and $\Psi_0=10.0$, is given in Fig.~\ref{fig:10}. It is noticed that the field values reached are almost two orders of magnitude higher ($1.26 \times 10^{14}$ G) than in the case of $\lambda=1$ (Fig.~\ref{fig:11}). Hence, the battery is now capable of twisting the poloidal field. We argue that the scenario in this simulation is quite realistic, as it does not require a high temperature along the crust, but only a region that may be greatly enhanced.


While it may be tempting to experiment more with variations of these parameters, or combinations of them, and how they may influence the architecture and evolution of the thermoelectrically induced magnetic field, we restrict ourselves to the experiments performed above. This is because the temperature distribution of neutron star has not been conclusively determined so far, and hence, any deep investigation of the chosen model parameters might be too theoretical, not resembling any particular relevance for similar future studies. The implications, practicability, and robustness of our model and results are discussed in the next section.

\section{Discussion} \label{sec:sec5} 

\subsection{Analysis} \label{sec:sec5.1}
\begin{figure*} 
\begin{tabular}{p{0.5\textwidth} p{0.49\textwidth}}
 a\includegraphics[width=0.4\textwidth]{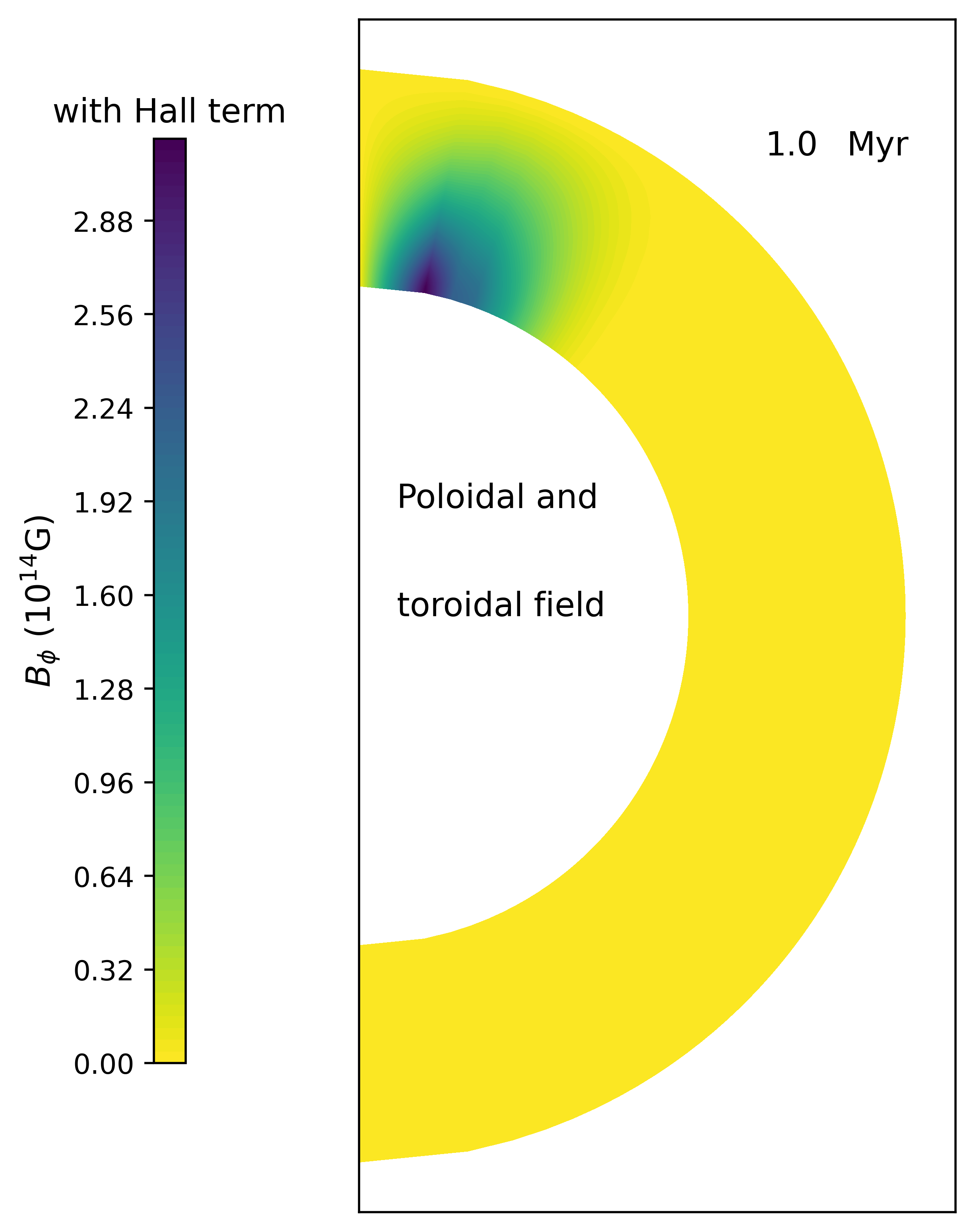}&
    b\includegraphics[width=0.4\textwidth]{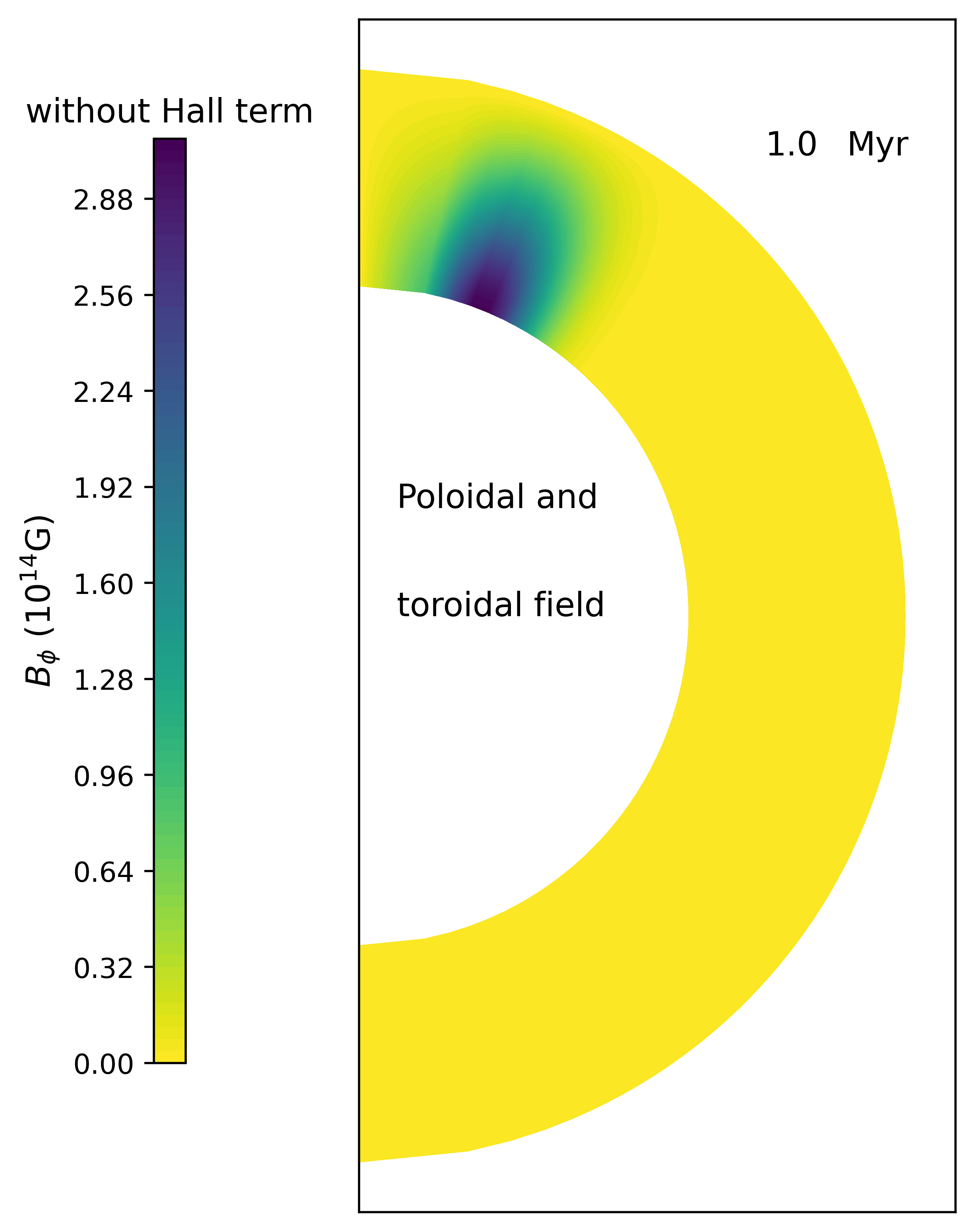}\\
     \end{tabular}
\caption{Final snapshots for Model A in the case of $\Psi_1$ and $\Psi_0 = 0$ with the Hall term (a) and without the Hall term (b).}
\label{fig:12}      
\end{figure*}

To further understand the thermoelectric mechanism in neutron star crusts, we now focus on simplified cases of our model and attempt to specify any general behavior of how the system evolves. We initially examine the case of a zero magnetic field as an initial condition, that is apart from $I_0=0$, we set $\Psi_0=0$ as well. Hence, we can isolate the influence of the poloidal field in the growth and resulting architecture of the toroidal one, as an initial purely toroidal field does not generate any poloidal field, whereas the poloidal field can generate a toroidal field. The final outcome (1 Myr) is given in Fig.~\ref{fig:12}a, for which all parameters other than $\Psi_0$ are kept the same as in Model A (Fig.~\ref{fig:2}). In both cases, the toroidal field grows linearly in time and saturates at the same values and timescales, as expected by eq.~(\ref{23}). However, in the absence of poloidal field, there is no field that can been twisted along the $\phi$-direction and generate Hall waves. Because of this, any toroidal field needs to be advected through the crust, and as this is a rather slow process, $\bm{B}_{T}$ remains zero far from the battery region instead of getting small negative values as occurs even with the lowest non-zero poloidal field.

We further simplify this scenario by omitting the Hall effects as well (Fig.~\ref{fig:12}b). The maximum strength of the toroidal field achieved is slightly lower in this case, $3.02 \times 10^{14}$ G, though its structure is modified. Compared to the case of including the Hall term in the induction equation (Figs.~\ref{fig:12}a or Fig.~\ref{fig:2}c), the region of the strong field appears to be more compact and shifted toward the equator. This behavior is of course expected, since, for a purely toroidal field, the Hall drift tends to move electrons along the lines of constant $\chi$ (\cite{Reisenegger:2007}, Fig. 1 therein), depending on the field's polarity. Consequently, when Hall effects are considered, the crust regions governed by strong fields move toward the pole and spread. We confidently determine that the high-$\bm{B}_{T}$ is transported toward the axis because of Hall drift, while the higher conductivity at larger depths causes the currents to move near the core, as an out-turn of Ohmic losses. Moving the gradient symmetrically in the southern hemisphere of the star, we obtain equivalent outcomes.

Another useful conclusion drawn by this simplified simulation is related to the comparative strengths of the involved terms in the induction equation. Specifically, although the Hall effect does play a role in determining the final structure and strength of the field, its contribution is rather small compared to the other two factors, and may be ignored as a first approximation. It does not affect the energy injected in the system and can only moderately modify the field configuration around the gradient region. As expected, the Hall effect is related with the transient behavior, which at sufficiently long time will fade away, whereas the final magnetic field geometry and intensity is primarily determined by the interplay between the battery term, which acts as a source, and the Ohmic diffusion.

The battle between the battery and Ohm dissipation reaches a steady state of equilibrium in all our simulations. 
The resulting field strengths and energy are dependent on the temperature and the temperature gradient. For instance, comparing Fig.~\ref{fig:11}a and Fig.~\ref{fig:11}b, one can see that the final toroidal field is four times higher when the maximum temperature is two times higher, while their profiles are the same. In other words, the final configuration of the crust's magnetic field is not insensitive to this input in particular. To elucidate, we have shown that for the gradients assumed in this work, the process becomes ineffective below $\sim$$10^7$ K. Apart from that, the poloidal energies (when including poloidal fields) dwindle exponentially in the course of time. This implies that the toroidal field is not capable of sustaining some of the poloidal field around it; the contribution of the poloidal field to the phenomenon is almost insignificant. All these remarks lead to the realization that the final toroidal field culminates a saturation equilibrium (under the proper conditions of activating the battery).

\begin{figure*}
\begin{tabular}{p{0.3\textwidth} p{0.3\textwidth} p{0.3\textwidth}}
  a\includegraphics[width=0.29\textwidth]{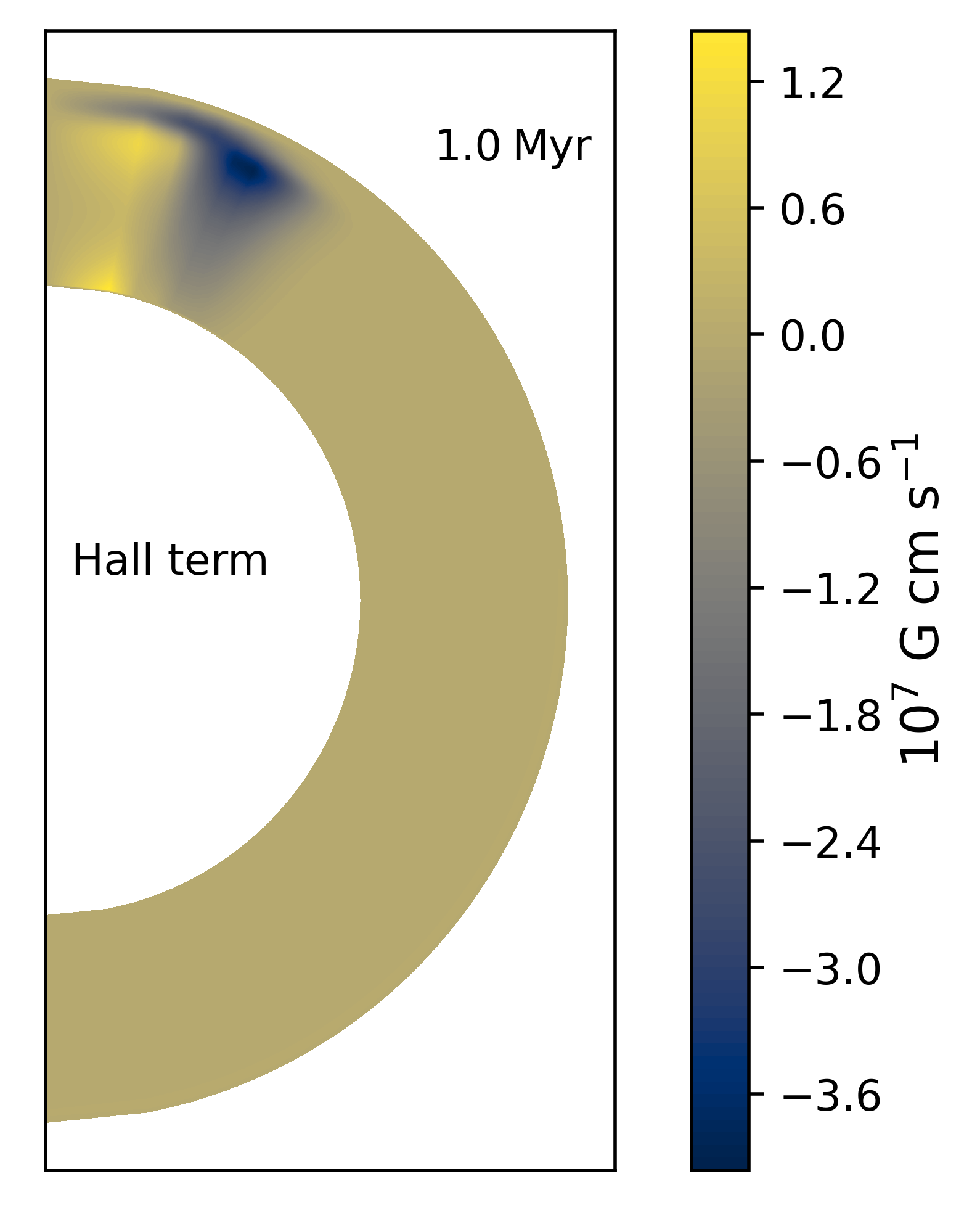}&
    b\includegraphics[width=0.29\textwidth]{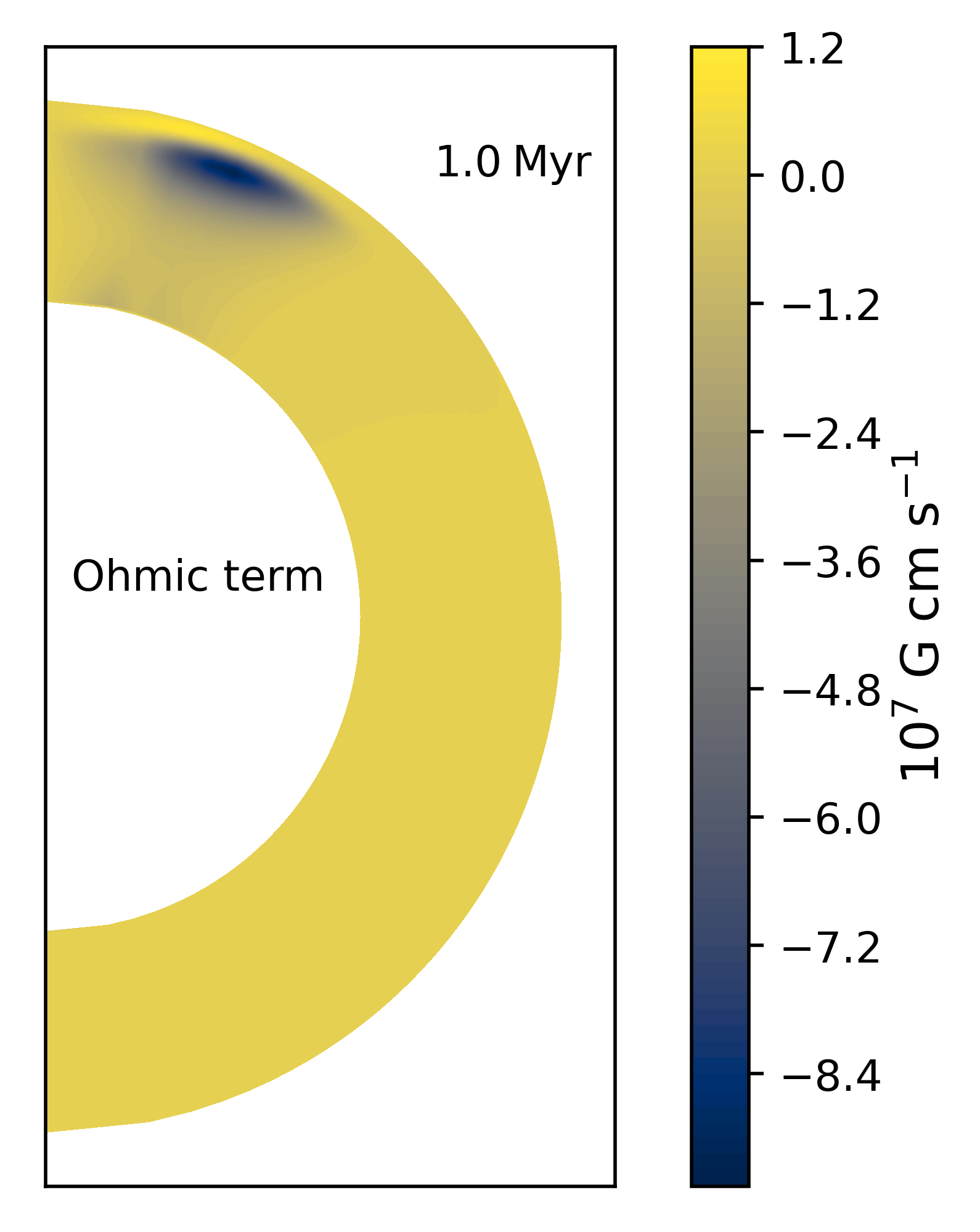}&
     c\includegraphics[width=0.29\textwidth]{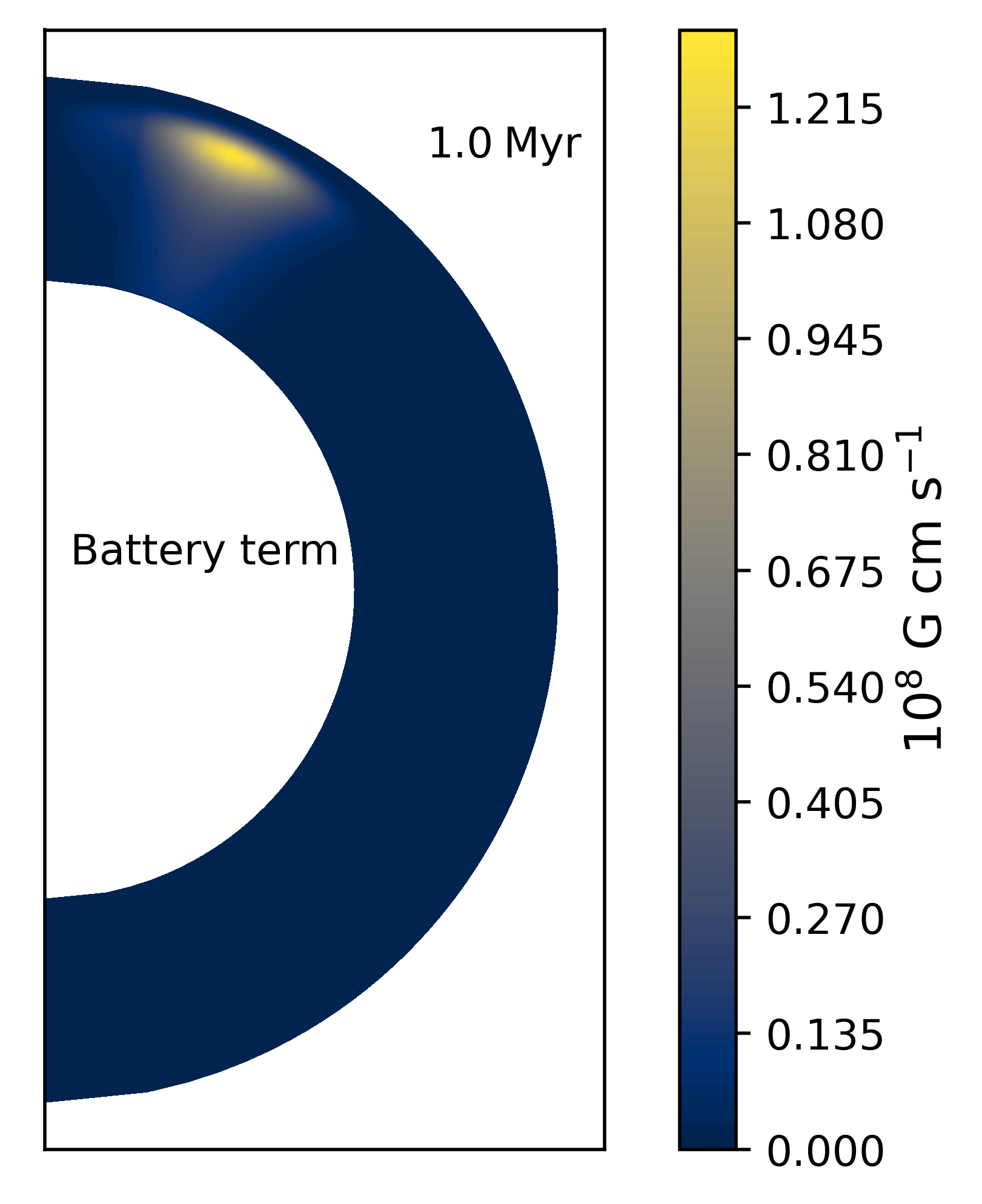} 
     \end{tabular}
\caption{Strength of the Hall (a), Ohmic (b) and battery (c) terms as deduced in eq.~(\ref{37}) for Model A in the case of $\Psi_1$ and $\Psi_0 = 0$ at 1 Myr (final step). The outermost radial grid layer of b panel (Ohmic term) was masked out for visualization purposes, due to the large values in this region ($\sim$6000).}
\label{fig:13}      
\end{figure*}

This saturation toroidal field can be approximately deduced semi-analytically. Based on eq.~(\ref{14}), we derive a scaling relation at saturation, $\partial I/\partial t = 0$. Ignoring the term involving the poloidal field ($\Psi$ does not play a particular role compared to $I$), the final expression writes as: 
\begin{eqnarray} 
I \left(\nabla \chi \times \nabla \phi \right) \cdot \nabla I  \nonumber\\ =\frac{c^{2}}{4 \pi \sigma}\left(\Delta^{*}I-\frac{1}{\sigma}\nabla I \cdot \nabla \sigma\right) + \frac{c}{e} r\sin\theta( \nabla S_e \times \nabla T ) \cdot \hat{\phi} \,.
\label{37}
\end{eqnarray}
The term on the left-hand-side corresponds to the Hall non-linear advection of the toroidal field \citep{2000PhRvE..61.4422V}, whereas the other two terms correspond to Ohmic losses and the battery. Fig.~\ref{fig:13}a-c show the strength of each term (Hall, Ohmic, and battery) throughout the crust at the final configuration step (1 Myr) for the model of Fig.~\ref{fig:12}a (Model A, $\Psi_1$, $\Psi_0 = 0$). In the high-$\bm{B}_{T}$ region, the battery term is dominant, reaching values around $1.3 \times 10^{8}$, and the Ohmic term is about $-9.5 \times 10^{7}$. They sum up to $3.6 \times 10^{7}$ representing the Hall term. The latter is three and two times weaker than the former terms, respectively (absolute values).

We can use dimensional analysis to estimate the scale of the saturation field, by keeping only the Ohmic and battery terms. The toroidal magnetic field obtained from $I$ writes as:
\begin{eqnarray}
B_{\phi}= \frac{I}{r \sin\theta} \sim \frac{S_e T L_S 4 \pi \sigma}{e c}= 
\frac{k_B^2 T^2 4\pi \sigma}{c^2 e \hbar }\left(\frac{\pi^4}{3 n_e}\right)^{1/3}\nonumber\\
=17B_{14}T_9^2\left(\frac{n_e}{10^{36}\ {\rm cm^{-3}}}\right)^{-1/3}\left(\frac{\sigma}{10^{24}\ {\rm s^{-1}}}\right)\,.
\end{eqnarray}
where $L_S$ is the scale at which the temperature and the magnetic field change, assuming that the battery term will activate currents within the region of the temperature anisotropy. 
The estimate of the magnetic field obtained through dimensional analysis is higher by a factor of 5 than the one typically found through simulations, and this can be attributed to the slight offset of the spatial structure of the magnetic field from that of the temperature. To quantify the impact of the Hall non-linear advection term we can use the results of the runs shown in Figs.~\ref{fig:12} a and b, where the impact of the Hall term affects the structure of the configuration, but not the maximum value of the toroidal field. In this sense, there is an analog of the Hall attractor, where the final outcome of the run converges to a magnetic field structure, where the toroidal field is determined by the interplay of the battery term, Ohmic dissipation and the Hall term. We obtain very similar results regardless of the initial field structure, and quite remarkably, even when no magnetic field is present in the initial conditions.

\subsection{Model practicability and limitations} \label{sec:sec5.2}

It is possible that temperature anisotropies are formed near the surface of a neutron star. Some possible energy sources that create a hotter region in a specific part of the crust are the following. First, it can be a result of the thermo-magnetic evolution of the star. For instance, the magnetic field itself might form hotspots through phenomena like reconnection. Additionally, the anisotropic heat conductivity due to the magnetic field can create horizontal temperature gradients, which is particularly interesting as it provides another coupling between the temperature and the magnetic field, so that there will be feedback in both directions. Second, it can be generated by backflowing magnetospheric currents. The current returning from the light cylinder forms a closed loop, and it may heat the region near the rim of the polar cap as it enters through the outer crust layers. Third, another possible mechanism only relevant to non-isolated neutron stars is accretion. Any accretion (e.g., through a disk) of different material (e.g., hydrogen) other than the iron-nickel onto the crust could lead to thermonuclear reactions, which increases the local temperature as well.

Whatever the mechanism is, it is believed that temperature differences may appear rather naturally on neutron stars. Numerous sources of strong magnetic fields and ages in the range of $10$ kyr have hotspots whose temperatures exceed $5\times 10^{6}$ K \citep{Vigano:2013}. For instance, in the magnificent seven, thermal pulses from kilometer-sized areas are observed \citep{Haberl:2007,Turolla:2009,Potekhin:2015,de-Grandis:2022}. These pulses correspond to temperatures of $10^6$ K while the characteristic ages of the sources are in the order of Myrs. Moreover, J0030+0451 has been thermally mapped in X-rays \citep{Riley:2019,Miller:2019}, and both circular and linear thermal regions of $\sim$$10^6$ K were identified near its south pole. While these temperatures are much smaller than those considered in this work, this source is rather old. Thus, surface temperature anisotropies can be maintained for rather long timescales. 

Thermal blanketing \citep{Gudmundsson:1983,Beznogov:2021} is another important point. The temperature in the interior of the crust can be much higher than the surface temperature, as also assumed by thermal evolution calculations \citep{Vigano:2013, de-Grandis:2020}. There, the surface  temperature is scaled to the crustal one by application of the following formula \citep{Igoshev:2021}:
\begin{eqnarray}
 \frac{T_c}{10^8 \ \mathrm{K}} = \left(\frac{T_s}{10^6 \ \mathrm{K}}\right)^2  \,,
\end{eqnarray}
where $T_c$ is the temperature at the top of the crust, and $T_s$ is the observed temperature at the surface. Thus, an observed temperature $5\times 10^6$ K, which is the case in several sources, corresponds to a crustal temperature of $2.5\times 10^9$ K, which lies well within the range of our calculations. We note though that the above relation of determining the inner crust temperature does not hold true when the heat comes from the surface hotspot, and it is only relevant for internal heat origin.

In our model, we assumed that temperature anisotropies appear in an idealized neutron star crust, ignoring their origin. Nevertheless, the main limitation of our considerations lies in the timescale on which this temperature gradient is maintained. We implicitly assumed that the thermal energy of the star rises in some way and remains constant over a period of hundreds of thousands of years. This implies that the temperatures we consider are somehow conserved by an external mechanism, although no such term exists in our equations (including the heat conduction in our set of equations is beyond the scopes of this paper). Realistically, temperature anisotropies drop, e.g. by neutrino emission \citep{Yakovlev:2001}, over the course of time (e.g., outburts at any point of the crust typically last for 5-10 years, or accretion may last for some thousand years). While we acknowledge this limitation, we think our study will be crucial first step in integrating battery terms in equations regarding the magnetic field evolution in neutron stars. Having calculated the maximum possible impact of a battery term, we show that at least for an idealized environment, thermoelectric effects are rather important and can potentially provide solutions to open problems regarding neutron star physics, hence thermo-magnetic interactions should not be omitted.

The suitability of the temperature function used was thoroughly reviewed in Section \ref{sec:sec3.2}. We now discuss the repercussions of the choices of its relevant parameters. We have assumed that the hot region of the star is merely connected to the polar cap. Its position in the meridional is given as a first approximation (for dipolar field):
\begin{equation}
    \theta_0 = sin^{-1} \left(\frac{r_{ns} \Omega}{c} \right),
\end{equation}
implying that it is related to the rotation of star ($r_{ns}$ stands for the neutron star radius and $\Omega$ is the rotational angular velocity). As we do not model a specific type of neutron star and restrict ourselves to an idealized situation, we did not underwent into much detail into the exact position of the central point, and placing it near the pole(s) will suffice. Other than that, the temperature gradient may be formed by outbursts as well, as explained, giving us the freedom to experiment with the geometry of the temperature profile. In this sense, that is why we examined a wide range for the extent of the temperature anisotropy.

Lastly, other minor limitations of our considerations are related to the models adopted for density and conductivity. Particularly, conductivity is a function of density, which in turn is dependent on the ocean temperature; thus, conductivity is that way only indirectly modulated by temperature (it also depends on temperature, at least as affected by phonon scattering). We have performed a test simulation assuming the conductivity linearly dependent on temperature, without noticing any significant changes in our outputs. Later work may adopt more accurate microphysics, including both electrical and thermal conductivities regulated by the respective temperature at each point.


\section{Conclusions} \label{sec:sec6}
Under appropriate conditions, magnetic field may be generated and enhanced through battery mechanisms, one of which is by temperature anisotropies causing thermal flux. High temperature gradients are required to achieve that, and neutron star crusts are a potential candidate for such a phenomenon. In this paper, we considered an idealized neutron star and showed that magnetic field enhancement might be possible through thermoelectric effects. 

In our model, we made the unrealistic approximation that a temperature profile is kept constant throughout the simulated time in order to assess the maximum effect the battery term has on the magnetic field evolution. Regions of localized strong toroidal field are formed (ranging from $10^{11}$ – $10^{15}$ G, depending on the temperature assumed), around of which poloidal field twists creating multipoles. For high temperatures ($\gtrsim 10^8$ K), the thermoelectric effect dominates the Hall drift, and the system requires some thousand years for Ohmic losses to be in equilibrium with the battery, with only minor oscillations observed afterward. Hall advection may only influence the final configuration of the toroidal field, rather than determine its value.

Later studies should shed more light into the actual role of thermoelectric effects in neutron stars. More realistic models need to be adopted, that is considering temperature decay and spread of heat through time or taking into account the thermo-magnetic interactions. Follow-up experiments with microphysical parameters and the temperature profile can be performed in the future as well, using the most up-to-date data deduced by observations. Lastly, we expect that 3-D simulations instead of axisymmetry might be worth investigating.

\begin{acknowledgements}
     We are grateful to an anonymous referee for constructive criticism that significantly improved the quality of this work. We thank Dimitris Ntotsikas for his assistance in the optimization of the code used for this work and Davide De Grandis for his insightful comments on the manuscript. KNG acknowledges discussions during the annual meeting of Simons Collaboration on Extreme Electrodynamics of Compact Sources, New York, 29 February to March 1st 2024. This work was supported by computational time granted from the National Infrastructures for Research and Technology S.A. (GRNET S.A.) in the National HPC facility - ARIS - under project ID pr015026/simnstar”.
\end{acknowledgements}


\bibliographystyle{aa} 
\bibliography{bibtex.bib} 

\end{document}